\documentclass[runningheads]{llncs}
\usepackage{pslatex}
\usepackage{amsmath}
\usepackage{xcolor}
\usepackage{newtxmath}
\usepackage{xspace}
\usepackage{paralist}
\usepackage{hyperref}
\usepackage[capitalise]{cleveref}
\usepackage{multirow}
\usepackage{wrapfig}

\usepackage{todonotes}

\newcommand{\toolname}[1]{\textsc{#1}\xspace}
\newcommand{\mona}{\toolname{Mona}}
\newcommand{\ours}{\toolname{eNfa}}
\newcommand{\cpp}{C++}
\newcommand{\vata}{\toolname{VATA}}
\newcommand{\brics}{\toolname{Brics}}
\newcommand{\dotnet}{\toolname{Automata}}
\newcommand{\zthree}{\toolname{Z3}}
\newcommand{\jaltimpact}{\toolname{JaltImpact}}
\newcommand{\abc}{\toolname{bwIC3}}
\newcommand{\minisat}{\toolname{Antisat}}
\newcommand{\bisim}{\toolname{Bisim}}
\newcommand{\cvc}{\toolname{CVC5}}

\newcommand{\bw}[1]{#1^{\mathsf{b}}}
\newcommand{\bwnfa}[1]{\bw #1}
\newcommand{\fw}[1]{#1^{\mathsf{f}}}
\newcommand{\fwnfa}[1]{\fw #1}
\newcommand{\ach}{\mathsf{Ach}}
\newcommand{\symbvar}{\alpha}
\newcommand{\pow}[1]{\mathcal{P}(#1)}
\newcommand{\benchname}[1]{\textsf{#1}\xspace}
\newcommand{\bcsmtlib}{\benchname{b-smt}}
\newcommand{\bregexlib}{\benchname{b-regex}}
\newcommand{\bhand}{\benchname{b-hand-made}}
\newcommand{\bparam}{\benchname{b-param}}
\newcommand{\bincl}{\benchname{b-armc-incl}}
\newcommand{\arandom}{\benchname{a-ltl-rand}}
\newcommand{\aparam}{\benchname{a-ltl-param}}
\newcommand{\aspec}{\benchname{a-ltlf-spec}}
\newcommand{\asloth}{\benchname{a-sloth}}
\newcommand{\aspagheti}{\benchname{a-noodler}}
\newcommand{\altlf}{\benchname{a-ltlf-patterns}}
\newcommand{\diffsat}{\benchname{diff-sat}}
\newcommand{\diffunsat}{\benchname{diff-unsat}}
\newcommand{\intersat}{\benchname{inter-sat}}
\newcommand{\interunsat}{\benchname{inter-unsat}}
\newcommand{\exppathsone}{\benchname{exppaths1}}

\usepackage{trimclip}

\makeatletter
\DeclareRobustCommand{\shortto}{%
  \mathrel{\mathpalette\short@to\relax}%
}

\DeclareRobustCommand{\shortminus}{%
  \mathrel{\mathpalette\short@minus\relax}%
}

\newcommand{\short@to}[2]{%
  \mkern2mu
  \clipbox{{.5\width} 0 0 0}{$\m@th#1\vphantom{+}{\rightarrow}$}%
}

\newcommand{\short@minus}[2]{%
  \mkern2mu
  \clipbox{{.5\width} 0 0 0}{$\m@th#1\vphantom{+}{-}$}%
}
\makeatother

\newcommand{\labeledto}[1]{{{\shortminus}\hspace{-2pt}\raisebox{0.16ex}{$\scriptstyle\{ #1\hspace{-0.28pt}\}$}\hspace{-2.2pt}{\shortto}}}

\newcommand{\scriptlabeledto}[1]{{{\shortminus}\hspace{-1.0pt}\raisebox{0.12ex}{$\scriptscriptstyle\{ #1\hspace{-0.28pt}\}$}\hspace{-1.6pt}{\shortto}}}

\newcommand\move[3]{%standard LaTex syntax where each parameter is put in {}
\mathchoice
{#1\,\labeledto{#2}\,#3}
{#1\labeledto{#2}#3}
{#1\scriptlabeledto{#2}#3}
{#1\scriptlabeledto{#2}#3}
}

\newcommand{\concat}{\cdot}
\newcommand{\concats}{\cdots}

\newcommand{\langof}[1]{\lang(#1)}
\newcommand{\lang}{L}

\newcommand{\regex}[1]{\texttt{#1}}
\newcommand{\ltlf}{$LTL_f$\xspace}

\newcommand{\M}{\mathcal{M}}

\newcommand{\boolform}[1]{\mathbb B(#1)}
\newcommand{\alphapred}{\mathbb{P}}

\newcommand{\conf}{c}

\newcommand{\aut}{\mathcal{A}}
\newcommand{\symbolf}{\varphi}
\newcommand{\statef}{\psi}

\newboolean{showcomments}
\setboolean{showcomments}{true}
\ifthenelse{\boolean{showcomments}}
{ \newcommand{\mynote}[3]{
    \fbox{\bfseries\sffamily\scriptsize#1}
    {\small$\ll$\textsf{\emph{\color{#3}{#2}}}$\gg$}}}
{ \newcommand{\mynote}[3]{}}
\newcommand{\shrink}[1]{}
\definecolor{pink}{rgb}{1,0.2,0.7}
\definecolor{purple}{rgb}{0.7,0,0.9}
\definecolor{darkgreen}{rgb}{0,0.5,0}
\begin{document}
%%\title{Comparing Solvers Regular Properties and Alternating Automata
%\title{Solvers of Boolean and Alternating Combinations of Regular Properties: Comparative Study
\title{Reasoning about Regular Properties:\\ A Comparative Study
\vspace{-1mm}
%\thanks{
%This work has been supported by 
%the Czech Ministry of Education, Youth and Sports ERC.CZ project LL1908, and
%the FIT BUT internal project FIT-S-20-6427.}
}

\author{
Tomáš Fiedor \and
Lukáš Holík \and
Martin Hruška \and\\
Adam Rogalewicz \and
Juraj Síč \and
Pavol Vargovčík
%\vspace{-1mm}
}

\institute{
Brno University of Technology\\
\email{\{ifiedortom,holik,ihruska,rogalew,sicjuraj,ivargovcik\}@fit.vutbr.cz}
}
\maketitle              
\vspace{-3mm}
\begin{abstract}
Several new algorithms for deciding emptiness of Boolean combinations of regular languages and of languages of alternating automata (AFA) have been proposed recently,
especially in the context of analysing regular expressions and in string constraint solving. 
The new algorithms demonstrated a significant potential, 
but they have never been systematically compared, 
neither among each other nor with the state-of-the art implementations of existing (non)deterministic automata-based methods. 
In this paper, we provide the first such comparison as well as an overview of the existing algorithms and their implementations.
We collect a diverse benchmark mostly originating in or related to practical problems from string constraint solving, analysing LTL properties, and regular model checking, and evaluate collected implementations on it.
The results reveal the best tools and hint on what the best algorithms and implementation techniques are. Roughly, although some advanced algorithms are fast, such as antichain algorithms and reductions to IC3/PDR, they are not as overwhelmingly dominant as sometimes presented and there is no clear winner. The simplest NFA-based technology may be actually the best choice, depending on the problem source and implementation style.  
Our findings should be highly relevant for development of these techniques as well as for related fields such as string constraint solving. 
%We also present an implementation of the antichain algorithm for emptiness test of alternating automata using a novel tight combination with a SAT solver as a means of symbolic handling of large alphabets, and show that it is competitive and sometimes superior superior with the state of the art.  
\vspace{-2mm}
\end{abstract}
%

%\enlargethispage{3mm}
\section{Introduction}
\vspace{-1mm}
Efficient representation of regular properties of finite words has been the subject of research for a long time, with applications and results spanning much of the field of formal reasoning, 
including regular expression matching, verification, testing, modelling, or general decision procedures of logics. 
% When regular properties are combined or generated using Boolean and similar operations, 
% interesting decision problems, including the most essential problem of language emptiness (further just emptiness), are PSPACE-complete. 
When regular properties are combined or generated using Boolean and similar operations, 
interesting decision problems are PSPACE-complete. 
This includes the most essential problem of language emptiness (further just emptiness),
The textbook approaches that use deterministic automata are in practice plagued by state space explosion. 
Determinisation and complementation is done by exponential subset construction and conjunction is quadratic. 
This motivated the research on efficient algorithms for non-deterministic 
and alternating automata. 
Using nondeterminism and alternation, one can gain one or two levels of exponential savings in the size of automata, respectively.  
Alternation, which adds conjunctive branching to the disjunctive non-deterministic branching, allows to avoid the blow-up in the automata size completely. However, from the perspective of the worst case complexity, the gained succinctness is payed back by the PSPACE-completeness of language emptiness. Still, the more succinct the representation gives more opportunities for clever heuristics that combat the worst case complexity and work in practical cases, essentially by avoiding re-creation of the entire (non)deterministic representation.  

Several very promising techniques and their implementations were proposed during the recent years.
%in the areas of analysing temporal properties,
%automata decision procedures for MSO and its fragments,
%processing of textual data: in analysing structured configuration files, regular properties used for filtering and sanitizing, and generally in analysing programs manipulating strings.
The latest advances in testing AFA emptiness appeared in the context of analysing
combinations of regular expressions and in string solving. 
A group of these
techniques is based on reducing AFA emptiness to a reachability in a
Boolean transition systems and using existing implementations of model-checking algorithms, most notably of IC3/PDR \cite{hoder_pdr_2012,bradley_safety_2007}, such as ABC \cite{abc}, nuXMV \cite{nuxmv}, or IC3Ref \cite{ic3ref},
to solve it
\cite{cox_paper_17,cox_mosca_19,janku_string_2018,fangyu_circuit_16}. 
The most recent contribution from \cite{margus_derivatives_21} extends the SMT-solver Z3 with symbolic derivatives, a generalisation of Antimirov derivatives of regular expressions. Z3 uses them to convert a combination of regular expressions into an alternating/Boolean automaton and on the fly tests its language emptiness through the classical de-alternation and a search for an accepting configuration.

Slightly older algorithm for testing equivalence of AFA (convertible to an emptiness test) is based on computing bisimulation up-to congruence \cite{dantoni_afa_2016}. It generalizes the original NFA-equivalence test of \cite{bonchi_checking_2013}.   
The congruence closure algorithms were preceded by the antichain algorithms that optimise the subset construction by the subsumption pruning \cite{dewulf_antichains_2006,ganty_fixed_2010}, and by the first attempt to use the model checking algorithms, namely the algorithm Impact of \cite{mcmillan_lazy_2006}, to emptiness of combinations of regular properties \cite{gange_unbounded_13}. 
%(intersection in particular).  
%
Lastly, the area of string constraint solving gave rise to a large variety of string constraint solvers.
%, that, despite targeting much more rich language than only regular properties. 
They approach combinations of regular properties through a spectrum of clever techniques based e.g. on automata, transformations to other types of constraints, reasoning on lengths of strings, Parikh images, etc. (e.g. Z3 \cite{z3,margus_derivatives_21}, CVC4/5 \cite{cvc5,cvc422}, Z3Str4 \cite{Z3str4}, OSTRICH \cite{AnthonyReplaceAll2018,AnthonyComplex2019}, Trau \cite{ChainFree,Trau} to name a few).
%(e.g. solvers like Z3, cvc4/5, Z3Str4/3RE, Trau, Norn \cite{cvc422,cvc5,BTV09,z3,Z3str4,Z3str3RE,Trau,AutomataSplitting} and dozens of others)

These works demonstrate a significant promise, 
but they are presented in specific, often narrow contexts and under varying views on state of the art. Consequently, they were never sufficiently compared against each other. 
%
%Generally, new results in the area are presented under wildly varying contexts and views on the state of the art, making the area difficult to navigate. 
%
Even comparisons against the most efficient implementations of the more standard techniques based on (non)deterministic automata is rare. 
String solvers were compared only against string solvers, advanced AFA-emptiness tests were compared only against the basic de-alternation.
A somewhat interesting comparison was done only between NFA-antichain and up-to congruence-based language inclusion and equivalence test in \cite{bonchi_checking_2013} and in \cite{chenfu_eqchecking_17}, and between the basic antichain based AFA emptiness and a version that uses abstract interpretation \cite{ganty_fixed_2010}. A number of works also take as their baseline implementations of automata 
%such as BRICS automata library \cite{brics} 
or string solvers 
%such as Norn \cite{norn}, 
which, even though being respectable tools in their own right, 
%which 
are currently not the fastest solvers of combinations of regular properties in either category.
On top of that, all the mentioned works on solving combinations of regular properties use only narrow benchmarks, often mutually exclusive.
%The lack of systematic comparison makes the landscape of these algorithms and the tools is thus difficult to navigate.
%

\enlargethispage{3mm}
Systematic comparisons of tools and algorithms on meaningful benchmarks is obviously 
needed to answer the questions `What to use?' and `What to compare with?',  
and generally for the field of reasoning about regular properties and automata to progress.
We thus present a comparison of implementations of major algorithms. %that we are aware of and able to run. 
%We also implement the AFA emptiness check of \cite{ganty_fixed_2010} that combines antichains with abstraction,  
%and our own version of the antichain algorithm of \cite{dewulf_antichains_2006} with a novel tight integration with a SAT-solver to handle large symbolic alphabets.
%
We compare the tools on a large benchmark of problems that we have collected from other works, 
from string constraint solving problems, analysis of regular expressions,  
regular model checking, and analysing LTL properties of systems.   
We believe that it is currently the most comprehensive benchmark in existence.
%
%The benchmark contains regular combinations that can be given to all the tools, and AFA emptiness problems, that can be given only to the tools handling general AFA. 
%
%Our examples do not cover the whole scale of applications of alternating automata and Boolean regular combinations. 
%
Our main focus is on examples around string solving and analysis of regular expressions, 
which is also where the most of the recent developments has happened. 
%
%These are problems that can be handled reasonably well using automata with explicit transition relation. That is, examples where it is not an absolute necessity to use complex and fine tuned symbolic representations of large numbers of transitions over many symbols, such as the bit-vectors that arise in solving WS1S or Presburger arithmetic. 
%
%
These benchmarks mostly allow for a relatively simple representations of automata transition functions. 
Even though the alphabets in examples coming form this are large (e.g. UNICODE with up to $2^{32}$ symbols), 
the alphabet size can, in most cases, be reduced to few symbols by working with alphabet minterms (classes of indistinguishable symbols) instead of individual symbols. 
% 
%which specifying classes of symbols on the input as intervals of numbers is linear to the input size (versus exponential for e.g. general bit-vector alphabets with character classes specified as formulae).  
%
The issue of effective symbolic representation of transition relations with large alphabets then does not dominate the evaluation, although it would be critical in other application areas, such as deciding WS1S or arithmetic formulae.  

We have obtained results that paint the basic landscape of the available techniques and tools. They identify tools and approaches which are likely to work well and should be used as the baseline in comparisons. 
We also provide a relatively diverse and large benchmark to be used in comparisons. 
The results broadly confirm that the new algorithms represent a leap in efficiency compared to the technology of DFA and also make a reduction of a problem to language emptiness of alternating automaton an attractive  option. 
On the other hand, they challenge some folklore knowledge and conclusions implied elsewhere.
For instance, reductions to IC3/PDR, although yielding one of the fastest algorithm, are not as vastly superior as sometimes presented.  
%
%There is in fact no clear winner. \todo{update}
%
Some practically relevant benchmark categories are best solved by a combination of an antichain algorithm with a SAT solver. Others, surprisingly many in fact, by a simple efficiency oriented implementation of basic algorithms for nondeterministic automata. 
%
%String constraint solvers are generally not the fastest on this kind of a problem, but sometimes use heuristics that solve examples than no other algorithm can. 
%
Our results also underscore that there is no universal silver bullet. 
The particular kind of the problem, determined to a large degree by its source, is a decisive factor that should be taken into account when choosing and tuning a solver.  

%
%it is still be advantageous in comparison to exponentially larger NFA as it gives opportunities for property driven search heuristics. 
%%
%Recently this has been recognised and utilised also by several work especially in the context of string solving,
%%
%where Boolean-like combiantions of regular properties arise naturally.
%%
%There has been a growing interest in emptiness testing of alternating automata, 
%especially in the constext string constraint solving. 
%\cite{}
%
%%
%The naive emptiness check namely uses de-alternation, a transformation to an exponentially large NFA, followed by the classical linear-time test of reachability of a final NFA state.
%%
%AFA however offer unique opportunities in practical use. 
%%
%In applications where the task is only to decide language emptiness, creating an explicit representation is not important,
%%
%the implicit AFA representation allows to take an advantage of the property driven approach:
%%
%The precise language, or its NFA/DFA representation, are not important and can be optimized or simplified by abstraction as long as the emptiness of the language stays preserved.
%%

\vspace{-3mm}
\section{Preliminaries}
\enlargethispage{4mm}
\vspace{-2mm}
\label{sec:preliminaries}

%\paragraph{Finite Automata.}
A~\emph{(nondeterministic) finite automaton (NFA)} over $\Sigma$ is a tuple
$\aut= (Q,\Delta,I,F)$
 where $Q$ is a finite set of \emph{states},
$\Delta$ is a set of \emph{transitions} of the form $\move q  a  r$ with $q,r\in
%Q$ and $ a \in\Sigma\cup\{\epsilon\}$, $I\subseteq Q$ is the set of \emph{initial states}, and $F\subseteq Q$
Q$ and $ a \in\Sigma$, $I\subseteq Q$ is the set of \emph{initial states}, and $F\subseteq Q$
is the set of \emph{final states}.
A~\emph{run} of~$\aut$ over a~word~$w \in \Sigma^*$ is
a~sequence
 $\move{p_0}{ a _1}{p_1}
  \move{}{ a _2}{}
  \ldots
 %$p_{n-1},  a _n, p_n$  
 \move{}{ a _n}{p_n}$  
where for all $1\leq i \leq n$, it holds that $ a _i \in \Sigma \cup
\{\epsilon\}$, $w =  a _1
\concat  a _2 \concats  a _n$, and either $\move{p_{i-1}} { a _i} {p_i}\in\Delta$ or $p_{i-1} = p_i$, $a_i=\epsilon$.
%  $p_0,  a _1, p_1 $, 
%  $ a _2, p_2 $,
% $\ldots$, 
%  %$p_{n-1},  a _n, p_n$  
%  $ a _n, p_n$  
% where $\move{p_{i-1}} { a _i} {p_i}\in\Delta$ if $1\leq i \leq n$ 
% is a \emph{run} of $\aut$  from $p_0$ to $p_n$ over the word $w =  a _1 \concat  a _2 \concats  a _n$ ($ a _i=\epsilon$ are neutral). 
The run is \emph{accepting} if $p_0 \in I$ and $p_n\in F$, and the \emph{language}
$\langof{\aut}$ of $\aut$ is the set of all words for which $\aut$ has an accepting run.

The automaton is \emph{deterministic (DFA)} if for every state $q$ and symbol $a$, $\delta$ has at most one transition $\move q a r$.
Any NFA can be determinised by the \emph{subset construction}, which creates the DFA
$A' = (2^Q,\Delta',\{I\},\{S\mid S\cap F\neq \emptyset\})$ where $\move S a S'\in \Delta'$ iff $S' = {\{s' \mid s\in S \land \move s a {s'}\in\delta\}}$.   
The basic automata constructions implementing Boolean operations with languages are
intersection, 
$\aut\cap\aut' = (Q\times Q',\Delta^\times,I\times I',F\times F')$ where 
$\move {(q,q')}{ a }{(r,r')}\in\Delta^{\times}$ iff 
$\move{q}{ a }{r}\in\Delta$ and 
$\move{q'}{ a }{r'}\in\Delta'$, non-deterministic union $\aut\cup\aut' = (Q\cup Q',\Delta\cup\Delta',I\cup I',F\cup F')$, deterministic union by product which is the same as $\cap$ up to that the final states are $F\times Q \cup Q\times F$, and complementation which consists of determinization and complementing the final states.

\vspace{-3mm}
\paragraph{Alternating automata.}
An \emph{alternating finite automaton (AFA)} in the most general form would be a tuple 
$\M = (\Sigma,\alphapred, Q,\delta,I,F)$ where, when denoting $\boolform X$ the Boolean predicate formulae over variables $X$:
\begin{inparaenum}[1)]
  \item $\Sigma$ is a finite alphabet;
  \item $\alphapred$ is a set of unary \emph{symbol predicates} with a free variable $\symbvar$;
  \item $Q$ is a finite set of \emph{states};
  \item $\delta: Q \rightarrow \boolform{Q \cup \alphapred}$ is a
    \emph{transition function} where states of $Q$ have only positive occurrences
  \item $I \in \boolform{Q}$ is a positive \emph{initial condition}; and
  \item $F\in\boolform{Q}$ is a negative \emph{final/accepting condition}.
\end{inparaenum}

It can be interpreted as the \emph{forward NFA} $\fwnfa A = (\Sigma, \pow Q,\fw \Delta, I',  F')$ with states $\conf\subseteq Q$ called \emph{configurations} of $A$.   
Assume many sorted interpretation of formulae over variables $Q$ of the type Boolean (values $0$ and $1$) and the variable $\alpha$ of the type $\Sigma$.
A set of states $\conf\subseteq Q$ is understood as an assignment $Q\rightarrow \{0,1\}$ in which $\conf(q)=1$ corresponds to $q\in \conf$.
A pair $(\conf,a)$, $a\in\Sigma$ is understood as the same assignment extended with $\alpha \mapsto a$. The satisfaction relation $\models$ between a formula and a configuration $\conf$ or a pair $(\conf,a)$ is defined as usual. 
The transition relation $\fw \Delta$ then contains a transition $\move \conf a {\conf'}$ iff 
$(\conf',a) \models \bigwedge_{q\in\conf} \Delta(q)$, and
$I'$ and $F'$ are the sets of configurations that satisfy $I$ and $F$, respectively.
It is common to define $\fw \Delta$ to contain only the smallest transitions, that is, for a given $\conf$ and $a$, 
only the transitions $\move \conf a {\conf'}$ with the $\subseteq$-minimal target $\conf'$ are in $\Delta$.\footnote{A state in a configuration is understood as a constraint. The less constraints, the more can be accepted from the configuration. Transitions to more constrained configurations are useless.} 
The language of $A$, $L(A)$, is the language of $\fwnfa A$.

The AFA can equivalently be interpreted as the \emph{backward NFA}, the automaton $\bwnfa A = (\Sigma, \pow Q,\bw \Delta, I', F')$ where
$\move \conf a {\conf'} \in \bw \Delta$ if  
$(\conf,a)\models \Delta(q)$ for each $q\in \conf$. 
Here it is enough to take, for a given $\conf'$ and $a$, only the transition with the $\subseteq$-largest source $\conf$\footnote{Going backward, larger configurations are more permissive. Transitions from the same target with smaller configurations are useless.} 
%The set of final configurations is also pruned from configurations included in other final configurations. } 
(this makes the transition relation backward deterministic).
%$\bw I$ equals $\fw I$ and the only final configuration in $\bw F$ is $Q \setminus F$. 
%A noteable property if this automaton is that it is backwards deterministic, yet still only singly exponential to the size of $A$.

%Here, we use $[\varphi]$ to denote the formula obtained from $\varphi$ by substituting every state $q$ by the primed version $q'$. 
%
%$\bigwedge_{q\in Q}q\rightarrow [\Delta(q)]'$. Here, we use $[\varphi]$ to denote the formula obtained from $\varphi$ by substituting every state $q$ by the primed version $q'$. 
%
%The successor of a configuration $c$, is a solution of 
%$\exists Q\exists \symbvar  \, \Phi_\Delta \land \bigwedge_{q\in C} q$. 

%Yet another definition generates a DFA with states from $\boolform(Q)$. 
%Its initial state is $I$, every formula $\varphi$ such that $\varphi\land F$ is satisfiable is a final state, 
%and $\move \varphi a {\varphi'}\in \Delta$ 

\enlargethispage{1mm}
\vspace{-2mm}
\paragraph{Boolean automata.}
Alternating automata may be extended to Boolean finite automata (BFA) by allowing any Boolean combination in the initial, final, and transition formulae
(states in the initial and transition formulae may occur negatively, states in the final formula may occur positively). 
%However, for simplicity of presentation, we will talk about AFA only.
Note that the extension of AFA to BFA is not dramatic, as a BFA is easily encoded as an AFA with only double the size, by the following steps:
1) for each $q\in Q$, add state $\bar q$ with $\Delta(\bar q) = \neg\Delta(q)$,
2) transform all formulas in $I,F,\Delta$ to NNF, 
3) replace all literals $\neg q$ by $\bar q$ in $\Delta$ and $I$ and replace literals $q$ by $\neg \bar q$ in $F$.

\vspace{-2mm}
\paragraph{Restricted forms of AFA transition relation.}
%Alternating automata from our benchmark are used in several different forms that can be seen as special cases of our definition. 

The general form of AFA, as defined above, is the most succinct. It provides space for most optimizations, such as in \cite{vargovcik_21}. Automata in this form are generated from LTL conversions of \cite{ltlf} used in \cite{dantoni_afa_2016,vargovcik_21}. On the other hand, only a small subset of algorithms and tools support AFA in this most liberal form. 
%, namely, only the algorithms based on reduction to transition system reachability and our symbolic antichain algorithm. 
%
A common restriction (used e.g. in \cite{dantoni_afa_2016}) is to separate symbols from states in the transition formulae, that is, having $\Delta(q)$ in the form  $\symbolf \land \statef$ with $\symbolf\in\boolform{\alphapred},\statef\in\boolform{Q}$. We call such AFA \emph{separated}. 
%
%This is the case of most applications of AFA, for instance the case of the symbolic AFA of \cite{dantoni_afa_2016}.
%
The transition relation can then be seen as a function $Q\rightarrow \boolform\alphapred \times \boolform Q$. 
Separated AFA are often considered with the state formula $\statef$ in the disjunctive normal form (e.g. in \cite{doyen-antichain-10,ganty_fixed_2010}), which we call the  \emph{DNF form}, and $\Delta$ then may be seen as a set of transitions of the form $\move q \symbolf c$ where $\bigwedge c$ is a (positive) clause of $\statef$. 
%The DNF form is easy to implement and practically efficient, unless in relatively rare cases when the transformation to DNF tends to explode.\todo{!?not true!? in our experiment?}

\vspace{-2mm}
\paragraph{The decision problems.}
We will concentrate on two decision problems:
%\begin{itemize}
%\item
\\(1) \emph{AFA emptiness} asks whether the language of the given AFA is empty.
%\item 
\\(2) \emph{Emptiness of Boolean combinations of regular properties} (\emph{BRE}), asks whether a Boolean combination of regular languages, given as automata or regular expressions, is empty (languages can be combined with $\cap$, $\cup$, and complement wrt. $\Sigma^*$, which also covers testing inclusion and equivalence\footnote{$L'\subseteq L$ is emptiness of $L'\cap \overline L$ and equivalence is emptiness of  $(L'\cap \overline L)\cup(\overline{L'}\cap L)$.}). 
%\end{itemize}

%\section{Tools for AFA and B-FA Testing}
\vspace{-2mm}
\section{Existing Algorithms and Tools}
\vspace{-1mm}
In this section, we will overview the existing approaches and tools implementing AFA and BRE emptiness.

%Boolean automata were experimented with in \cite{cox_paper_17,cox_mosca_19} as means of testing combinations of regular expressions, and in a sense also in \cite{fangyu_circuit_16} in the context of solving string constraints. 
%Non-determinisation of BFA is also a part of \cite{margus_rex_10}.  
%

\vspace{-3mm}
\subsection{Representation of Automata Transition Relations}
\vspace{-1mm}
%An important parameter of the implementations is the encoding of transitions.  
%The generic predicates in the definition above cover essentially all cases appearing in the literature. 
%(alphabet predicates have been long used in many applications of automata and were most systematically defined and treated in the works on symbolic automata \cite{}).
In the simplest form, a predicate on a automata transition represents a single letter from the alphabet. 
This is called an \emph{explicit transition}. Explicit automata are simple, allow for low level optimizations, and implementation of complex algorithms for them is manageable (such as advanced algorithms for computing simulations \cite{ranzato_efficient_2010,lukasjirisimulation,Cece17}). 
%For instance, transitions may be ordered by symbol, which makes various synchronous exporations of automata fast.  
The technique of a-priori mintermization, that replaces the alphabet by the alphabet of minterms, classes of indistinguishable symbols, makes explicit automata usable also when alphabets are large. 
%They work well in our benchmarks from regular model checking, from string solving, and from processing regular expressions. 
However, when the number of minterms tends to explode, explicit automata do not scale. 
%

%\todo{
%This happens e.g. in solving WS1S \cite{Buchi62} 
%or Presburger arithmetic with automata \cite{BoudetC96,WolperB95}, in certain approaches to string constraint solving \cite{janku_string_2018}, processing LTL formulae with an alphabet  of sets of atomic propositions. 
%}

%Our becnhamrks from regular model checking are in the explicit form and the automata we obtain from mintermisation, with the alphabet of minterms, are also using explicit transition function. 

Various implementations of automata have been using transition predicates implemented as BDDs, Boolean formulae, formulae over SMT-theory of bit-vectors, intervals of numbers, etc.
This has been systematized in the works on \emph{symbolic automata} \cite{margus_rex_10,dantoni_taminimization_2016,margus_the_power17}, where the symbol predicates may be taken from any effective Boolean algebra (and the automata are in the separated form).  
Even more compact than symbolic automata are representations of the transition relation used in the WS1S solver \mona or in some of the implementations of AFA, which in a way drop the restriction to the separated form. 
We will discuss the concrete implementations below. %in the next section.

\paragraph{}

\enlargethispage{5mm}
\vspace{-12mm}
\subsection{(Non)deterministic Finite Automata}
\vspace{-1mm}
The baseline approach to solve BRE is to use DFA or NFA. 
Boolean operations are implemented as the classical construction listed in \cref{sec:preliminaries}.
Automata may be kept deterministic, 
%in which case deterministic union by product construction may be used 
%(used in MONA) or automata are always determinised (in BRICS). 
or they are kept non-deterministic whenever possible and determinised only before complementing. 
%(in our implementation, BRICS, VATA, AutomataDotNet). 
An important ingredient of achieving efficiency is usually to minimize automata at least once every few operations (important e.g. in applications such as regular model checking \cite{bouajjani-antichain-08} or some approaches to string solving \cite{noodler23,AnthonyReplaceAll2018,Trau}).
The deterministic approaches construct the minimal DFA by the Hopcroft, Moore, Brzozowski, or the Huffman algorithm \cite{hopcroft_71,Moore1956,Brzozowski1962CanonicalRE,HUFFMAN1954}, 
the non-deterministic approach may use simulation \cite{ranzato_efficient_2010,Cece17,HHK95,Ilie2004,lukasjirisimulation} or bisimulation \cite{Valmari_10,piagetarjan_87,symbsim18} based reduction methods.
Simulation reduces significantly more but is much costlier.
DFA/NFA are implemented in many libraries. Here we select a representative sample. 

First, \ours is the simplest tool, our own implementation of NFA. It uses explicit automata with mintermization. It is implemented in \cpp, with efficiency in mind, but with no extensive optimizations (roughly, transitions from a state stored in a two layered data structure, the first layer divided and ordered by symbols, and the second layer ordered by the target state). It uses 
an off the shelf implementation of one of the newest generation algorithms for computing simulation \cite{ranzato_efficient_2010,Cece17,lukasjirisimulation} (that achieve good efficiency through a usage of the partition-relation data structure) taken from \vata tree automata library \cite{vata} (implementing namely \cite{lukasjirisimulation}).\footnote{In our experiment, simulation is only used after parsing and has minimal overall impact.}

The \brics automata library \cite{brics}, written in Java, is often considered a baseline in comparisons \cite{brics}. It uses primarily deterministic automata and symbolic representation of transition relation using character ranges. It is written in Java and is relatively optimized. 

The \dotnet library  \cite{automatadotnet}, made in C\#, implements symbolic NFA/DFA  parame\-trized by an effective Boolean algebra. We use it with the default algebra of BDDs. \dotnet has been long developed and has accumulated many optimizations and novel techniques for handling symbolic automata \mbox{(e.g., optimized minimization \cite{margus_minimization}).} 
%or \cite{margus_rex_10}).

%Symbolic NFA parametrizable  by an effective Boolean algebra were implemented in \cite{automatadotnet,margus_minimization}, 
%AFA in \cite{dantoni_afa_2016} and in \cite{margus_derivatives_21} (as symbolic Boolean derivatives). \cite{margus_minimization} shows that symbolic automata open space for optimizations not available to explicit representations, that may play a role already on alphabets of a moderate size. 

\mona \cite{mona}, written in C, is the most influential and optimized implementation of deterministic automata. 
It specialises in deciding WS1S formulae, which besides Boolean combinations includes also quantification. The decision procedure generates DFA with complex transition relations over large alphabets of bit-vectors. For this purpose, \mona uses a~compact representation of the transition relation: a single MTBDD for all transitions originating in a state, with the target states in its leaves. \mona can represent only a~DFA, hence it always implicitly determinizes. 

\vata \cite{vata}, written in \cpp, is a library implementing non-deterministic tree automata. 
As NFA are a special case of tree automata, we can use it as an implementation of the basic constructions for explicit NFA. It is relatively optimised. We include it into the comparison for its fast implementation of the antichain inclusion checking \cite{bouajjani-antichain-08,tree_inclusion_11}, which for NFA boils down to the inclusion check of \cite{doyen-antichain-10}.

\vspace{-4mm}
\subsection{Alternating Automata}
\vspace{-1mm}

\paragraph{De-alternation.}
The basic approach to AFA emptiness is \emph{de-alternation}, transformation to an NFA, either the forward $\fwnfa A$ or the backward $\bwnfa A$, followed by testing the emptiness of the resulting NFA. 
Both NFAs are constructed by a variation on the NFA subset construction. 
We are not aware of any tool using pure de-alternation,
and we believe that it would not be competitive. 
The forward algorithm is however the basis of \cite{margus-derivatives-21} used in \zthree where it is run on the fly with a novel symbolic derivative construction (discussed also in the paragraph on string constraint solvers).

\vspace{-2mm}
\paragraph{Interpolation based abstraction refinement.}
Attempts to harness model checking algorithms to AFA emptiness appeared in the context of string solving and processing of regular expressions.
To our best knowledge, the earliest attempt was \cite{gange_unbounded_13}, where conjunctions of regular constraints were solved using the interpolation-based algorithm of \cite{macmillan-interpolation-03}.
The interpolation-based abstraction refinement, namely the algorithm Impact of \cite{mcmillan_lazy_2006}, was also used in \cite{radu18}. This work concentrated on more general problem, solving emptiness of AFA over data words with an infinite data domain (that can relate past and current values of data variables). Their tool \jaltimpact \cite{jaltimpact} (in Java), that we include into our comparison, can be run on our benchmark too. 
%In the light of it being written in Java and not primarily meant for our kind of problem, it show a good potential. 

%\paragraph{IC3/PDR}
\vspace{-2mm}
\paragraph{Reduction to reachability and IC3/PDR.}
The work of \cite{fangyu_circuit_16} presented the first translation of string constraints (mostly BRE) into reachability in a Boolean transition system (circuit) that was then solved  by the model checker nuXmv \cite{nuxmv}. This was de facto the first reduction of AFA emptiness to reachability in a Boolean transition system (BTS). 

Let us briefly overview the basic principle of the reduction.  
The \emph{forward BTS} for an AFA $A$ has configurations that are Boolean assignments to $Q$, 
initial and final configurations satisfy $I$ and $F$, respectively,
and transitions are given by the formula 
$\fw \Phi_\Delta : \bigwedge_{q\in Q}q\rightarrow [\Delta(q)]'$. 
Here we use $[\varphi]'$ to denote the formula obtained from $\varphi$ by substituting every state $q$ by its primed version $q'$,
and we will also denote by $[\conf]'$ the primed version $\{q' \mid q \in \conf\}$ of a configuration $\conf$. 
A \emph{successor} of a configuration $\conf$ is any configuration $\bar \conf$ such that 
$[\bar \conf]'$ satisfies 
$\exists Q\exists \symbvar  \, \fw \Phi_\Delta \land \bigwedge_{q\in C} q$ (the symbol variable alpha is of the bit-vector sort).
\emph{Reachability} is then the transitive and reflexive closure of the successor relation and the \emph{reachability problem} asks whether a final configuration is reachable from an initial one. It is the case if and only if $A$ is not empty. 
\cite{fangyu_circuit_16} used the forward reduction. %\todo{fakt?}
Alternatively, the \emph{backward BTS} for $A$ has 
the initial configurations satisfying $F$, final configurations satisfying $I$, and the successor relation given by the formula
$\bw \Phi_\Delta : \bigwedge_{q\in Q}q'\rightarrow \Delta(q)$.

Works of \cite{cox_paper_17,cox_mosca_19} applied IC3/PDR \cite{hoder_pdr_2012,bradley_safety_2007}, implemented in IC3Ref \cite{ic3ref}, together with the backward BTS reduction to solve emptiness of BRE and obtained very encouraging results.  
Their implementation is, however, proprietary and not publicly available.
Similar approach was taken by \cite{janku_string_2018}, where a string constraint was translated to a multi-tape AFA and then to a BTS by the forward translation, and given to IC3/PDR to solve through tools nuXmv~\cite{nuxmv} or ABC~\cite{abc}. 
%
%
%were so called automata splitting \cite{AutomataSplitting}, that reduces membership terms of the form $x.y\in L(A)$ into a large Boolean combiantion of purely regular constaints, was succinclty represented by alternating automata and given to IC3/PDR to solve through tools nuXmv or ABC \cite{nuxmv,abc}.
%
%All these work essentially transform an AFA into a Boolean transition system and call a model model checker to decide its emptiness. 
%\paragraph{AFA as transitions systems.}
%Some, by a reduction used some of the best tools in our comparison.
%
Results of \cite{vargovcik_21} seem to indicate that the backward translation is better and the same is suggested by the comparison in \cite{cox_paper_17,cox_mosca_19} in which the string solver Sloth \cite{janku_string_2018}, based on the forward reduction, was much slower that Qzy, based on the backward reduction. In this comparison, we include our own \cpp{} implementation \abc of the backward reduction based on \mbox{the model checker ABC.}

%
%The works \cite{janku_string_2018,fangyu_circuit_16} encode the transition systems that encodes the NFA constructed by the forward subset construction, 
%\cite{cox_paper_17,cox_mosca_19} encodes the backward construction. 
%%
%In \cite{}, we conjectured that the forward backward reachability is more efficient. This is also indicated by the comparison with \cite{janku_string_2018} in \cite{cox_paper_17,cox_mosca_19}. In this work, we include into the comparison our own implementation of the backward approach, that ultimately uses the tool ABC to decide the PDR problem. 

%IC3/PDR then searches throucgh the search-specae of these NFA while utilising a form of property driven abstraction refinement.\todo{discuss this}

\enlargethispage{3mm}
\vspace{-2mm}
\paragraph{Antichains.}
Antichain algorithms presented in \cite{dewulf_antichains_2006} were the first breakthrough in solving BRE. They use subsumption relations between the states of the automata constructed by variations of the subset construction to prune the constructions. 
They were used to test language universality and inclusion of NFAs and AFA emptiness.
The AFA emptiness namely is based on an on-the-fly search for an accepting state of the $\fwnfa A$ or for an initial state of the $\bwnfa A$. 
Subsumption prunes discovered states that are larger (smaller for the backward algorithm) than others. 

The antichain algorithms were enhanced and generalized in a number of works, e.g. with a more aggressive pruning by the simulation-based subsumption \cite{abdulla_when_2010,doyen-antichain-10}, or by counterexamples guided abstraction refinement in \cite{ganty_fixed_2010}.
%
%As noted also in \cite{chenfu_eqchecking_17}, 
%which compared antichain NFA inclusion of  \cite{} and the bisimulation up-to congruence algorithm of \cite{}, complex optimisations such as simulation subsumptoion often do not pay off in practical cases. 
%also come with a cost and pay off only when implemented carefully. In fact, it seems that simple implementations are superior in most cases. 
%Since we are not aware of such optimised implementations that could hadle our benchmark, 
%
In this comparison, we include the NFA inclusion check implemented in the \vata tree automata library
%\cite{holik-tainclusion-11,bouajjani-antichain-08,vata}, 
\cite{vata}.
We also experimented with a student-made implementation of the antichain AFA emptiness check of \cite{ganty_fixed_2010}  that uses abstraction refinement (the original implementation is no longer maintained and we were not able to run it). However, not being  able to achieve a competitive performance, we excluded it from the comparison. One reason of the poor performance may be that simplest form of AFA, explicit DNF form (used in the original version \cite{ganty_fixed_2010}), might be too inefficient and costly to construct in our examples, partly due to a large number of minterms induced by the AFA emptiness benchmark. 

%Our AFA antichain check that integrated with a SAT solver seems to work particularly well. 
We implemented (in \cpp) the antichain AFA emptiness test of \cite{doyen-antichain-10} that integrates tightly with a SAT solver to handle the general form of AFA with large alphabets. We will refer to it as \minisat.
We will briefly explain its principle.
It essentially implements the reachability test for the backward BTS discussed in the previous paragraph. 
A configuration $\conf$ is represented by the conjunction $\phi_\conf = \bigwedge_{q \in Q \setminus \conf} \neg q$.
Note that $\phi_\conf$ is satisfied by the downward closure of $\conf$, which are all configurations included in (subsumed by) $\conf$.
To compute predecessors of configurations represented by $\phi_\conf$,
the SAT solver (namely MiniSAT \cite{minisat03}) is called on the formula 
$\Phi: \bw \Psi_\Delta \land \phi_\conf \land \psi_{\ach}$. 
Here, $\psi_{\ach}$ excludes all already discovered configurations from the solution. It is a conjunction of clauses $\overline{\phi_\conf} : \bigvee_{q\in Q\setminus \conf} q$ for every previously discovered configuration $\conf$.  
The SAT solver discovers a satisfying assignment $e$, which is turned into a new configuration $c' = Q \cap e$ (that is, the values of the symbol bits constituting the bit-vector $\symbvar$ are omitted from $e$). Unless $c'$ is initial, it is queued for further predecessor computation and is immediately added to $\phi_\ach$ through the interface of incremental SAT solving as the clause $\overline{\phi_{c'}}$.
Finally, only maximal predecessors of $c$ are of interest, as the non-maximal ones are subsumed by them. We enforce the maximality of $c$ through working directly with the internal SAT solver structures: at decision points, the SAT solver is forced to give priority to decisions that assign 1 to state variables.

%
%The VATA tree automata library can be with NFA, and contains a fast implementation of %antichaian based NFA inclusion checking and of simulation reduction. 
%

%\cite{,fiedor_nested_2015,fiedor_lazy_2017,dreytel_coalgebras_15} (more precisely, implicit subsumption between NFA states created by classical NFA constructions or AFA configurations), 
%\todo{pavol antichain}

\vspace{-2mm}
\paragraph{Bisimulation up-to congruence.}

A later class of algorithms, here refered to as \emph{up-to algorithms}, checks equivalence as a bisimulation between configurations of AFA, and utilises the up-to congruence technique to prune the search space.
The first algorithm on NFA equivalence \cite{bonchi_checking_2013} was extended to alternating automata emptiness check in \cite{dantoni_afa_2016}.
These algorithms are close to antichains. \cite{bonchi_checking_2013} shows that the pruning potential of the up-to techniques is in theory the same or larger than that of antichain.
A disadvantage of the up-to congruence technique is the need for expensive evaluation of congruence closures.
The more extensive experiments of \cite{chenfu_eqchecking_17} shows antichain algorithms as faster, with an exception of randomly generated automata with small alphabets and very dense transition relations.
We include into the comparison the Java implementation of the AFA-emptiness of \cite{dantoni_afa_2016} (emptiness reduces to equivalence with a trivial empty AFA), that we refer to as \bisim. The other implementations of up-to algorithms we are aware of, from \cite{chenfu_equchecking_17} and \cite{bonchi_checking_2013}, are single-purpose programs that decide equivalence of two NFAs, hence we would be able to run them on a very small fraction of \mbox{our benchmark only}. 
%\todo{in fact we should take the EBEC \cite{chenfu_equchecking_17}}.

%\paragraph{String constraints solvers.}
\vspace{-3mm}
\subsection{String Constraints Solvers}
\vspace{-2mm}
There are dozens of string constraint solvers that implement, to a various degree, a support for deciding combinations of regular properties. String languages are rich and BRE are not the absolute priority of the solvers, hence they perform on them generally worse than specialised tools. However, string solvers implement a wide scale of unique techniques and pragmatic heuristics that may work in specific instances. Representatives of the solvers with the most mature implementations (also used in most comparisons in the literature) are \zthree \cite{z3,margus-derivatives-21} and \cvc \cite{cvc422,cvc5}. \cvc solves BRE mostly through rewriting rules. 
%, Z3str3RE \cite{Z3str3RE} that uses automata-based length aware heuristics. 
Recently \cite{margus-derivatives-21} extended \zthree with an approach based on the Antimirov derivative automata construction generalised to symbolic automata and extended regular expressions. Essentially, the construction produces a symbolic AFA/BFA and checks its emptiness on the fly while running the forward de-alternation.
\cite{margus-derivatives-21} shows that it is significantly more efficient in solving BRE than other SMT solvers (including \cvc).   

%\paragraph{Other approaches and tools.}
\vspace{-3mm}
\subsection{Other Approaches and Tools}
\vspace{-2mm}
Although we believe that we have collected a representative subset of existing algorithms and tools, we have not collected all interesting specimens. Some were not available, some were difficult to run or prepare the inputs for, some seemed covered by experimentation in other works. Including these tools and algorithms into the comparison could still be interesting and we leave it for the future work (we plan to keep extending the tool base as well as the benchmark set). Namely, the tool DPRLE \cite{dprle}, used in the comparison in \cite{cox_paper_17}, seemed to be mostly outperformed by the IC3/PDR approach implemented in Qzy, however, not absolutely consistently. The implementation of NFA antichain and up-to congruence techniques used in \cite{chenfu_eqchecking_17} seems efficient, with its NFA antichain inclusion twice as fast as that of \vata. The up-to congruence NFA equivalence checking of \cite{bonchi_checking_2013} could be fast too (\cite{bonchi_checking_2013} and \cite{chenfu_eqchecking_17} report somewhat conflicting results). There are numerous NFA/DFA libraries, e.g. the C alternative of \brics \cite{libfa} or the Java implementation of symbolic NFA of \cite{lorisjava}. ALASKA \cite{alaska} might contain interesting implementations of antichain algorithms but is no longer maintained and available. Our comparison is missing a basic implementation of antichain-powered de-alternation for explicit AFA in the DNF form, which, if not overwhelmed by a large number of minterms, could reach a good performance through simple fast data structures, similarly to our \ours.

\vspace{-3mm}
\section{Benchmarks}
\vspace{-1mm}

We collected as comprehensive benchmark as possible, harvesting examples used in previous works as well as generating some of our own. It is available at \cite{experimentweb}.
Our main focus is the areas where the most of the development in solving AFA and BRE emptiness happened recently, which is string constraint solving and analysis of regular expressions used in analysing and filtering texts. 
Atomic regular properties are here mostly given in the form of regular expressions over UNICODE character classes. The alphabet is large but the number of minterms is mostly small or moderate. 
This is true also for our examples from regular model checking. 
Symbolic handling of complex transition relations over large alphabets is thus not absolutely crucial and the experiment can stay focused on the main algorithms for emptiness check. 
For that reason, we do not include benchmarks from solving WS1S \cite{Buchi62}, the primary target of \mona, 
or Presburger arithmetic with automata \cite{BoudetC96,WolperB95}, where the techniques of handling symbolic alphabet are indispensable. 
Techniques specialising at this kind of problems would deserve their own study.  
Our benchmarks where the symbolic alphabet representation is still rather important are AFA coming from (combinations of) LTL properties, with alphabets of sets of atomic propositions, and from translations of string constraint problems to AFA with complex multi-track alphabets.%
\footnote{We did not attempt to generate purely random problems. First, purely random automata generated e.g. by \cite{tabakov-vardi_05} seem to have different characteristics than automata coming from practical problems (e.g. in \cite{bouajjani-antichain-08,chenfu_eqchecking_17}). Second, although generating random NFA is possible with a generator controlled by three simple parameters which give a manageable parameter-value space covering all NFA, it is not clear how to similarly generate random AFA or BRE. On the other hand, we do include a benchmark based on randomly generated LTL formulae, which we consider relatively close to realistic LTL specifications.}

\vspace{-2mm}
\paragraph*{Boolean combinations of regular expressions.}
This group of BRE contains benchmarks on which we can run all tools, including those based on NFA and DFA. They have small to moderate numbers of minterms (about 30 in average, at most over a hundred).
%
%
%first set of benchmarks contains Boolean formulas of regular membership constraints of the form $x \in L$ where $x$ is the only variable and $L$ is some regular language given by a regular expression.
%These formulas can be succinctly transformed into the problem of AFA emptiness, each constraint $x \in L$ is transformed into NFA for $L$ and Boolean operator is transformed into operator of AFA (with negation working as complement).
%Here we collected benchmarks from multiple sources:

\vspace{-1mm}
\begin{itemize}
\item[\bcsmtlib]
%inspired by~\cite{margus_derivatives_21},
contains 330 string constraints from the Norn and SyGuS-qgen, collected in SMT-LIB benchmark \cite{BarFT-SMTLIB}, that fall in BRE. These were also used to compare SMT-solvers in \cite{margus_derivatives_21}. 
% took from the set of benchmarks for string solving as collected by SMT-LIB~\cite{BarFT-SMTLIB} a subset of 323 %problems containing only constraints of type $x \in L$ for one variable $x$ which are combined by some Boolean operators, %(these were used in \cite{margus_derivatives_21}, but they could use more, because they implemented their method in general string solver Z3),
\item[\bhand] has 56 difficult handwritten problems from~\cite{margus_derivatives_21} containing membership in regular expressions extended with intersection and complement. They encode
(1) date and password problems,
(2) problems where Boolean operations interact with concatenation and iteration, and
(3) problems with exponential determinization.
%we have a set of problems  string solving (explain that not all, that only if multiple $x \in L$ are used and only if nothing else is used), RegExLib (multiple of the form $x \in L_1 \land x \in L_2$ where $L_i$ is given by a regex from RegExLib database + $L_1 \not\subseteq L_2$ which is transformed into $x \in L_1 \land \neg x \in L_2$), multiple handwritten stuff (regular expressions extended with intersection and complement, we only take those that do not have intersection and complement too deep in the regex, i.e only the outer level)
\item [\bincl]
contains 171 language inclusion problems from runs of abstract regular model checking tools (verification of the bakery algorithm, bubble sort, and a pro\-duc\-er-con\-sumer system) of \cite{bouajjani-antichain-08}. These examples were used also in \cite{chenfu_eqchecking_17,bonchi_checking_2013}. 
\item[\bregexlib] contains 500 problems, obtained analogously as in \cite{dantoni_afa_2016,vargovcik_21}, of the form $r_1 \land r_2 \land r_3 \land r_4 = r_1 \land r_2 \land r_3 \land r_4 \land r_5$, where each $r_i$ is one of the 75 regexes\footnote{\url{https://github.com/lorisdanto/symbolicautomata/blob/master/benchmarks/src/main/java/regexconverter/pattern\%4075.txt}} from RegExLib~\cite{regexlib} selected so that $r_1 \land r_2 \land r_3 \land r_4 \land r_5$ is not empty.
This benchmark is inspired by spam filtering, where we want to test whether a new filter $r_5$ adds anything to existing filters. 
We transformed this problem into the inclusion $r_5 \subseteq r_1 \land r_2 \land r_3 \land r_4$, and kept the original form for \bisim which expects an equivalence.
%(we transform the equality into formula $(R \land \neg L) \lor (\neg R \land L)$ where $R$ is the right side and $L$ the left side of the equality),\todo{maybe we should do $R \subseteq L$ and for D'Antoni bisim alg we should do directly equality} 
%something with mail, that we do $x \in L_1 \land x \in L_2 \land x \in L_3 \land x \in L_4 = x \in L_1 \land x \in L_2 \land x \in L_3 \land x \in L_4 \land x \in L_5$ and $x \in L_1 \land x \in L_2 \land x \in L_3 \land x \in L_4 = x \in L_1 \land x \in L_2 \land x \in L_3 \land x \in L_4 \land x \in L_4$ where we know that $x \in L_1 \land x \in L_2 \land x \in L_3 \land x \in L_4$ is sat or something (d'Antoni says that, not sure if true for us)
\item[\bparam] has 8 parametric problems. Four are from~\cite{gange_unbounded_13}: 
\\(1) $\regex{[a-c]a[a-c]}\{n+1\} \cap \regex{[a-c]a[a-c]}\{n\}$ (long strings),
\\(2) $\bigcap_{i=1}^n\regex{([0-1]}\{i-1\}\regex{0[0-1]}\{n-1\}\regex{0[0-1]}\{n-i\}\alpha_i\regex{)|([0-1]}\{i-1\}\regex{1[0-1]}\{n-1\}\regex{1[0-1]}\{n-i\}\alpha_i\regex{)}$ (exponential branching),
\\(3) $\bigcap_{i=1}^n\regex{.*(.}\{p_{10+i}\}\regex{)+}\alpha_i$ (exponential paths 1), and
\\(4) $\bigcap_{i=1}^n\regex{.+}\alpha_i\regex{0(.}\{p_{10+i}\}\regex{)+}$ (exponential paths 2), where $\alpha_1,\ldots,\alpha_n$ are disjoint character classes and $p_j$ is the $j$-th prime number.
%(1) is actually from~\cite{rex}. 
Another four are from \cite{cox_paper_17}:\\
(5)~$\verb|^.[01]*.1.[01]{n}.$| \setminus \verb|^.[01]*.0.[01]{n-1}.$|$ (sat. difference),\\
(6)~$\verb|^.[01]*.1.1.[01]{n}.$| \setminus \verb|^.[01]*.0.[01]{n+1}.$|$ (unsat. difference),\\
(7)~$\verb|^.[01]*.1.[01]{n}.$| \cap \verb|^.[01]*.0.[01]{n-1}.$|$ (sat. intersection) and\\ 
(8)~$\verb|^.[01]*.1.[01]{n}.$| \cap \verb|^.[01]*.0.[01]{n}.$|$ (unsat. intersection). 
    %$\regex{}$ multiple artificial shit + some set of 10 (actually 14) regexes from \cite{tenregexes} which are actually from RegExLib (they have some special properties) also used in some other papers (these benchmarks are probably shit, regexlib might be even subsumed by the benchmarks of \cite{margus_derivatives_21}),
%    one parametric problem from . 
For (1) we chose $n \in \{50,100,\dots,500\}$, for (2)-(4) we chose $n \in \{2,3,\dots,60\}$ and for (5)-(8) we chose $n \in \{50,100,\dots,100\}$.
\end{itemize}

%\vspace{-3mm}
\subsection{AFA benchmark}
%\enlargethispage{3mm}
%\vspace{-2mm}
The second group of examples contains AFA not easily convertible to BRE. Here we can run only tools that handle general AFA emptiness. Some of these benchmarks also have large sets of minterms (easily reaching to thousands) and complex formulae in the AFA transition function, hence converting them to restricted forms such such as separated DNF or explicit may be very costly. This also seems to be the main reason for which our implementation of \cite{ganty_fixed_2010} could not compete.
\begin{itemize}
\item[\altlf] comes from transformation of linear temporal logic formulae over finite traces (\ltlf) to AFA~\cite{ltlf}. The 1699 formulae are from \cite{ltlf:multiplebenchmarks}\footnote{\url{https://drive.google.com/file/d/1eOYGvm3C8sQ-9iyfZ8qx42K54hgrFNTC}} and they represent common \ltlf patterns which can be divided into two groups: (1) 7 parametric patterns (100 each) and (2) randomly generated conjunctions of simpler \ltlf patterns (999 formulae).
\item[\arandom] contains 300 \ltlf formulae obtained with the random generator of \cite{vargovcik_21}. The generator traverses the syntactic tree of the LTL grammar, and is controlled by the number of variables, probabilities of connectives, maximum depth, and average depth. 
We have set the parameters empirically in a way likely to generate examples difficult for the compared  solvers (the formulae have 6 atomic propositions and maximum depth 16).
\item[\aparam] has a pair of hand-made parametric \ltlf formulae (160 formulae each) used in~\cite{dantoni_afa_2016,vargovcik_21}: \emph{Lift}~\cite{ltl:lift} describes a simple lift operating on a parametric number of floors and \emph{Counter}~\cite{ltl:counter} describes a counter incremented modulo the parameter.
% (both these formulae are not in \ltlf, satisfiable in their original LTL semantics).
%
\item[\aspec] \cite{ltlf:multiplebenchmarks} contains 62 \ltlf formulae that specify realistic systems, used by Boeing~\cite{ltl:boeing} and NASA~\cite{ltl:nasa}.
The formulae represent specifications used for designing Boeing AIR 6110 wheelbraking system and for designing NASA NextGen air traffic control (ATC) system.
\item[\asloth] 4062 AFA emptiness problems to which the string solver Sloth reduced string constraints \cite{janku_string_2018}. The AFA have complex multi-track transitions encoding Boolean opetarations and transductions, and a special kind of synchronization of traces requiring complex initial and final conditions.
\item[\aspagheti] 13840 AFA emptiness problems that correspond to certain sub-problems solved within the string solver Noodler in~\cite{noodler23}. The AFA were created similarly as those of \asloth, but encode a different particular set of  operations over different input automata. 
\end{itemize}

\vspace{-3mm}
\section{The Comparison}
\vspace{-1mm}
\enlargethispage{3mm}
\label{sec:comparison}

We ran our experiments on Debian GNU/Linux 11, with Intel Core 3.4GHz
processor, 8 CPU cores, and 20 GB RAM. All experiments were run with the timeout of 60 seconds (increasing the timeout did not have a significant impact). Additional details as well as the virtual machine with the entire benchmark are available at \cite{experimentweb}.
%We ran the experiments in parallel
%(utilizing 6 cores) with timeout set to 60 seconds\footnote{We tried higher
%timeout, which, however, had no significant impact.}.

\vspace{-1mm}
\paragraph{Benchmarking infrastructure.}
The initial difficulty is that the tools expect different input formats and forms of automata and the benchmarks come in different formats as well. 
%With few exceptions, we were able to feed most of examples to all tools that can principially handle them. 
We converted all benchmarks to our internal AFA format, from which we generated formats supported by the AFA handling tools \jaltimpact, \abc, \minisat, and \bisim, or we extend the tools with a parser. 
The BRE benchmarks come from various sources. We first convert them into a master file which specifies the Boolean combination of atomic  NFA, 
each atomic NFA stored in a separate file. 
The SMT-lib format is generated for $\zthree$ and $\cvc$.
In the case of $\bhand$, $\bparam$, and $\bcsmtlib$, 
the atomic automata are translated from regular expressions using the parser of \brics, while in the case of \bregexlib, where the regexes contain features not supported by \brics, we use the parser from \bisim.
$\bcsmtlib$ and $\bhand$ requires first translating from SMT-lib to a regular expression. 
In the case of \bincl, the atomic automata come directly as NFAs, and 
are converted into formats of the individual BRE solvers (we again wrote parsers for some of the solvers), and to our AFA format for the AFA solvers. Every BRE solver was extended by an interpreter of the master file that reads the NFA/DFA from the generated solver-specific files (except the SMT solvers, which read SMT-lib).     
We note that due to some difficulties with internal structures, 
we currently cannot run \brics on \bincl, and due to the lack of a converter frm complex regular expressions and from pure NFA to the SMT format, we do not run \zthree and \cvc on \bregexlib and on \bincl.%
%(we however still reproduce the results of \cite{margus_derivatives_21} for \zthree and \cvc).
%\footnote{These issues are solvable with a nontrivial engineering effort. Running these tools on the said benchmark does not seem critical for the overall outcome. }

%

%\abc is a~command line application that generates from an AFA an AIGER \cite{} file for ABC \cite{}. 
%an input an~AFA automaton and a~program with automata operations that it
%interprets. 

%\bisim and \minisat are compiled applications that takes automata
%in cap'nproto format and tests for emptiness. \jaltimpact is a~Java application
%that takes AFA automaton in its own format. For each of the five libraries
%(\brics, \dotnet, \mona, \ours, \vata) wrote small program that takes as an
%input program that performs set of automata operations (intersection,
%complementation, etc.) on input automata in formats corresponding to each tool.
%Finally the SMT-solvers (\zthree, \cvc) take as an input smt formula in smtlib
%format.

\vspace{-1mm}
\paragraph{Measured data.}
We will present the results obtained  with BRE (where we run all the tools) and with AFA emptiness (where we run \abc, \minisat, \bisim, and \jaltimpact) separately. 
We also separate the results on examples from applications from results on parametric hand-made examples. 

\begin{figure}[h!]
  \centering
  \includegraphics[width=\textwidth]{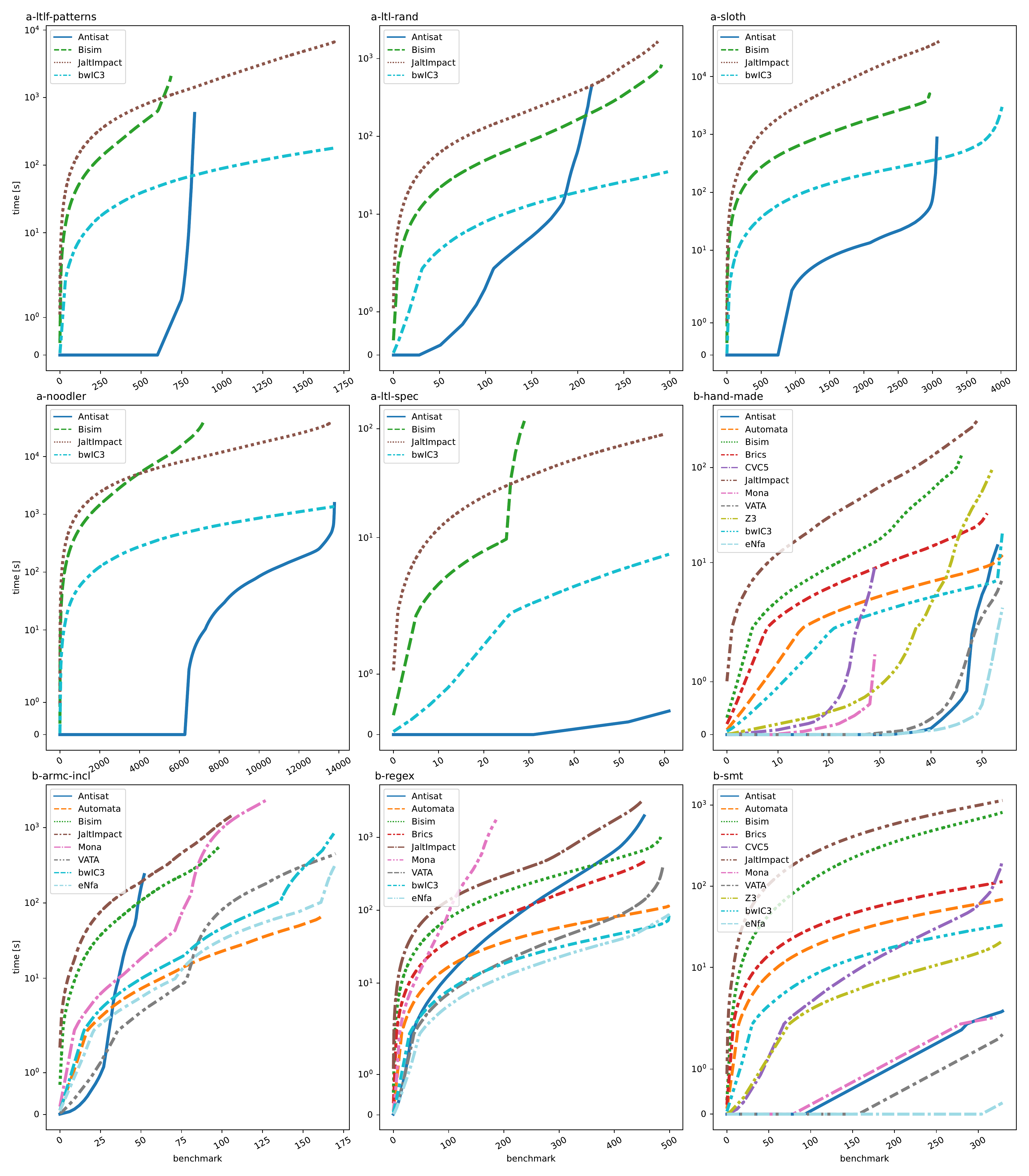}
        \vspace{-10.0mm}
  \caption{Cactus plots of AFA and BRE benchmarks. The $y$ axis is the cumulative time taken on the benchmark in logarithmic scale, benchmark on the $x$ axis are ordered by the runtime of each tool.}
  \label{fig:cactus}
\end{figure}

%Tables~\ref{tab:a} and \ref{tab:b} summarize 
Table~\ref{tab:b} summarizes
the statistics from evaluating the benchmarks. 
%Both tables list three metrics
The table lists: (i) the average time, (ii) the median time, and (iii) the number
of timeouts and number of errors (mostly, a tool ran out of the memory, made a~bad alloc or ran into a~segmentation fault). 
A few errors, e.g. in \cvc or \bisim, were due to the
unsupported features in the inputs.
The tools' performance is then visualised on cactus plots in  
\cref{fig:cactus}.
For each tool, the plot shows the progress of the tool on each
benchmark: the $y$ axis is the cumulative time taken on the benchmark, with the individual examples on the $x$ axis ordered by the runtime taken by the tool. Timeouts are omitted.
In the appendix, we also show a set of scatter-plots that compare for every benchmark the three best performing tools.
%from the graph. We scaled the $y$ axis to better highlight the faster tools.
%Graphs show, that many tools can complete the whole benchmark quite easily,
%while others struggle or even cannot complete the whole benchmark.

\newcommand{\win}[1]{\textbf{#1}}

\begin{table}[t]
\scriptsize
\centering
        %\vspace{-6.0mm}
\begin{tabular}{l|rrr|rrr|rrr|rrr|rrr|rrr}
\hline
 tool        & \multicolumn{3}{c|}{a-ltl-rand}   & \multicolumn{3}{c|}{a-ltl-spec}   & \multicolumn{3}{c|}{a-ltlf-patterns}   & \multicolumn{3}{c|}{a-noodler}   & \multicolumn{3}{c|}{a-sloth}         & \multicolumn{3}{c}{a-ltl-param}    \\
\hline
 \abc        & \win{0.1} & \win{0.1} & \win{0}              & {0.1} & {0.1} & {0}    & \win{0.1} & \win{0.1} & \win{0}      & \win{0.1} & \win{0.1} & \win{3}  & \win{1.3} & \win{0.1} & \win{34}  & \win{25.4} & \win{0.6} & \win{134} \\
 \bisim      & {4.4} & {1.0} & {8}              & 32.9 & 60.0 & 32                 & 37.0 & 60.0 & 1013                    & 31.6 & 26.4 & 6644(8)           & 17.5 & {1.5} & 1087(10)             & 58.2 & 60.0 & 308                  \\
 \jaltimpact & {7.9} & {2.3} & 12               & {2.4} & {1.4} & 0(1)             & {4.0} & {2.8} & {0}                   & {3.8} & {1.8} & 186             & 24.1 & 15.4 & 958                   & 47.0 & 60.0 & 205                  \\
 \minisat     & 18.3 & {0.1} & 84                & \win{0.0} & \win{0.0} & \win{0}              & 31.0 & 60.0 & 868                     & {0.4} & {0.0} & 57              & 14.9 & {0.0} & 991     & 58.3 & 60.0 & 310                  \\
\hline

\end{tabular}
        % \caption{Summary of AFA benchmarks. Table list (i) the average, (ii) the median, and (iii) the number of timeouts and errors (in brackets). Winners are highlighted in bold.}\label{tab:a}
        %\vspace{-10.0mm}
% \end{table}

% \begin{table}
% \scriptsize
% \centering
        %\vspace{-6.0mm}
\begin{tabular}{l|rrr|rrr|rrr|rrr|rrr}
\hline
 tool        & \multicolumn{3}{c|}{b-armc-incl}   & \multicolumn{3}{c|}{b-hand-made}   & \multicolumn{3}{c|}{b-regex}   & \multicolumn{3}{c|}{b-smt}   & \multicolumn{3}{c}{b-param}   \\
\hline
 \abc        & {5.2} & {1.1} & {1}               & {0.4} & {0.1} & {0}               & \win{0.2} & \win{0.1} & \win{0}           & {0.1} & {0.1} & {0}         & 44.9 & 60.0 & 191             \\
 \bisim      & 28.5 & {9.5} & 72                 & 11.2 & {1.0} & {8}                & {3.8} & {1.3} & 15            & {2.5} & {2.5} & {0}         & 55.4 & 60.0 & 240             \\
 \brics      & \multicolumn{3}{c|}{-}            & {3.9} & {0.4} & {3}               & {5.8} & {0.8} & 40            & {0.3} & {0.3} & {0}         & 52.7 & 60.0 & 228             \\
 \cvc        & \multicolumn{3}{c|}{-}            & 27.4 & {0.8} & 10(15)             & \multicolumn{3}{c|}{-}        & {0.8} & {0.2} & {1}         & 48.6 & 60.0 & 208             \\
 \dotnet     & {3.5} & {0.4} & {9}               & {0.2} & {0.2} & {0}               & {0.2} & {0.2} & {0}           & {0.2} & {0.2} & {0}         & 46.3 & 60.0 & 161(42)         \\
 \jaltimpact & 30.9 & 24.6 & 63                  & 11.1 & {3.6} & {5}                & 12.2 & {2.4} & 48             & {3.5} & {3.5} & {0}         & 57.8 & 60.0 & 252             \\
 \minisat    & 42.8 & 60.0 & 118                 & {1.4} & {0.0} & {1}               & {9.3} & {1.4} & 45            & {0.0} & {0.0} & {0}         & 39.0 & 60.0 & 147             \\
 \mona       & 28.5 & 44.1 & 43                  & 27.3 & {0.1} & 22(3)              & 41.0 & 60.0 & 15(298)         & {1.5} & {0.0} & {8}         & 44.9 & 60.0 & 25(169)         \\
 \ours       & \win{1.9} & \win{0.8} & \win{0}   & \win{0.1} & \win{0.0} & \win{0}   & \win{0.2} & \win{0.1} & \win{0} & \win{0.0} & \win{0.0} & \win{0}         & 44.6 & 60.0 & 143(51)         \\
 \vata       & {2.6} & {3.4} & {0}               & \win{0.1} & \win{0.0} & \win{0}   & {2.1} & {0.2} & 10(1)         & \win{0.0} & \win{0.0} & \win{0}         & 37.8 & 60.0 & 155(1)          \\
 \zthree     & \multicolumn{3}{c|}{-}            & {3.9} & {0.0} & {2}               & \multicolumn{3}{c|}{-}        & {0.4} & {0.0} & {2}         & \win{32.0} & \win{48.1} & \win{129} \\
\hline

\end{tabular}
%\caption{Summary of BRE benchmarks. Table list (i) the average, (ii) the median, and (iii) the number of timeouts and errors (in brackets). Winners are highlighted in bold.}\label{tab:b}
\caption{Summary of AFA and BRE benchmarks. Table list (i) the average, (ii) the median, and (iii) the number of timeouts and errors (in brackets). Winners are highlighted in bold.}\label{tab:b}
        \vspace{-3.0mm}
\end{table}

\begin{figure}[h!]
  \centering
  \includegraphics[width=\textwidth]{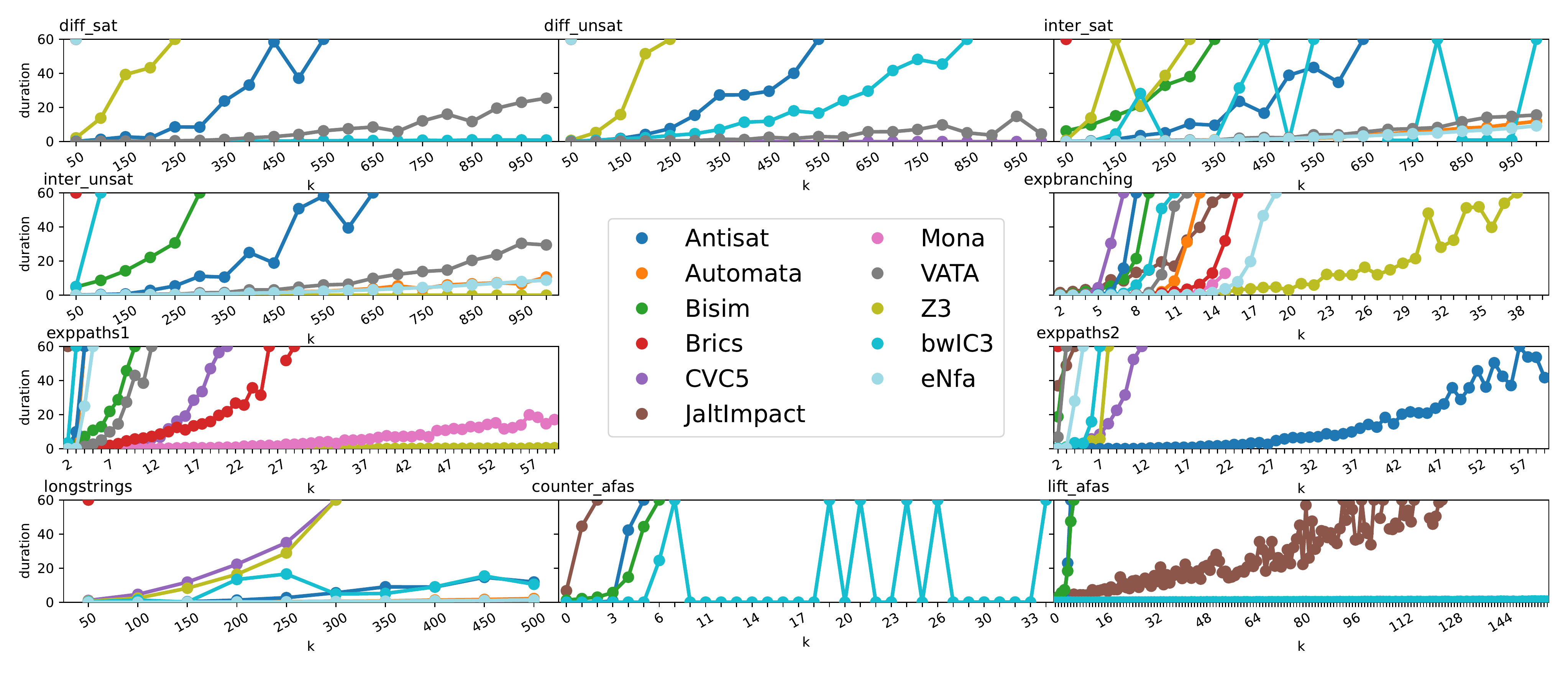}
        \vspace{-10.0mm}
    \caption{Models of runtime on parametric benchmarks based on specific parameter $k$ with timeout 60s. The sawtooths represent the tool failed on the benchmark for some $k$ while solving benchmarks for $k-1$ and $k+1$. For brevity, we draw the models only until they start continually failing.}
  \label{fig:models}
\vspace{-3mm}
\end{figure}

Finally, we compared the tools on the parametric benchmarks \aparam and
\bparam. We illustrate the results in Fig.~\ref{fig:models}. Each graph shows
the times for the increasing value of the specific parameter on the $x$ axis.

\vspace{-2mm}
\subsection{Discussion}
\vspace{-1mm}

Based on the measurements, we make several observations. 

Firstly, the tool which combines universality (it can be run on AFA as well as on BRE emptiness) with the most consistent good performance is \abc. It dominates most of the AFA emptiness benchmark, shows great or a very good performance on the BRE benchmark, and often stands out on the parametric examples. Moreover, the measurements reported in \cite{cox_paper_17} suggest that the backward BTS reduction has even more potential. This is visible namely from the comparison of our results on the parametric benchmarks \diffsat, \diffunsat, \intersat, and \interunsat.
Our implementation matched the result of \cite{cox_paper_17} on \diffsat and partially on \intersat, saw a worse trend on \diffunsat and much worse trend on \interunsat.
A likely culprit is a different underlying model-checker, ABC \cite{abc} in our implementation versus ICRRef \cite{ic3ref} in \cite{cox_paper_17}. 
\cite{cox_paper_17} however did not use IC3Ref out of the box, \mbox{harnessing IC3Ref efficiently is not entirely trivial.}  

Secondly, the results on application related BRE (all BRE except the parametric examples in \bparam) quite surprisingly favour the tools based mostly on relatively basic NFA algorithms. The overall best is the simplest tool of all, our implementation  \ours of basic NFA constructions. It should be emphasised that, despite being made with efficiency in mind, it is not optimised, and its simplicity allows for much further optimization. Close to the performance of \ours is \vata, which uses the antichain inclusion checking on \bincl and \bregexlib (the fact that explicit complementation of \ours is faster than the antichain of \vata suggests that the inclusion benchmarks are not particularly hard). \vata specialises to the more general tree automata, which probably causes unnecessary overhead. \dotnet also performs well. It uses slightly more advanced algorithms than \ours (such as lazy evaluation of difference, though, without antichain pruning). Its symbolic representation of transition functions with BDDs probably does not provide much advantage here. 
This result challenges the view that translating complex problems, arising for instance in string constraint solving, into AFA in order to use the sophisticated machinery of AFA solvers is an obvious silver bullet. Organizing the computation into smaller NFA operations, where, moreover, partial results can be minimised and re-used, and a simpler and hence more flexible NFA technology is used, might be a better strategy (this seems to work very well for instance in our recent prototype string constraint solver \cite{noodler23}).

%We don't know where the gap in efficiency comes from, although the difference in languages (\cpp vs C\#) and the overhead of the symbolic representation seem possible. 

Our AFA emptiness test \minisat based on the antichain algorithm and a SAT solver has an interesting performance. As can be seen on the cactus plots, besides its absolute domination on \aspec, it is significantly faster than other tools on a large portion of the other AFA emptiness benchmark, but struggles on the rest. The examples where it dominates are often 
%``flageral automata'' \cite{palodipl} 
automata with the structure resembling a lasso (or several lassos) with a long handle. The other implementation of an antichain algorithm, NFA/NTA inclusion in \vata, also shows a good performance. 
This together points on the overall strength of antichain algorithms.    

The SMT string constraint solvers are not among the best in the benchmark related to practical applications, but are competitive (especially \zthree), and win on some parametric cases. This may be due to that various heuristics unique to SMT solvers, especially rewriting that reduces one type of a constraint to another, kicks in. For instance, \zthree seems to solve
\exppathsone with a help of rewriting to the sub-string constraint in the theory of sequences. %\todo{check!!!} 
In general, the measurements on parametric examples underscore the fact that no algorithm is universally the best and their relative performance may vary drastically depending on the kind of an input.

Although the mediocre performance of the other tools can be partially explained by their focus on a different kind of a problem or a dated underlying technology, and each of them is respectable in its own right, a point can be made against relying on them as a baseline in comparisons of tools for solving our kind of problem. 
\mona, optimised for a different settings (complex alphabets of bit-vectors with many minterms), is held back by the implicit determinisation, and, in our case, probably by the overhead of the symbolic representation. It also frequently runs out of the 32-bit address space for BDD nodes. Similarly for \brics, which also always determinises.  
The low performance of \bisim is surprising relative to the good results of the up-to algorithms reported in \cite{bonchi_checking_2013,dantoni_afa_2016}. It is more consistent with \cite{chenfu_eqchecking_17} where up-to algorithms were not wining against antichains on the more practical examples. Our results however do not directly contradict the results of \cite{dantoni_afa_2016} itself, since it does not compare with the fast tools identified here and stands to a large degree on parametric and random benchmarks. 
%Although we are not aware of anything like that, 
There is also always the possibility that we have prepared the input in a way not ideal for the tool. For instance, transformation to the separated AFA, required by \bisim, is not entirely trivial. Further investigation of this and a comparison with some other implementation of the up-to techniques seems to be needed. The lack of a raw speed of \jaltimpact on BRE and AFA emptiness is expectable considering that it is meant for a different kind of systems, AFA over data words. The stable trends shown in the graphs suggest that an implementation of an interpolation-based abstraction refinement optimised for BRE and AFA emptiness might have a potential.   

\paragraph{Main takeaways.}
The backward reduction of AFA emptiness to BTS reachability in a combination with IC3 is very fast and extremely versatile, showing very good performance on almost all benchmarks. However, on BRE with a relation to a real world application, simple NFA algorithms actually tend to have the best raw performance, with the simplest implementation of NFA being the best. 
Antichain algorithms work also well, even significantly better than other algorithms on specific kinds of AFA. 
These seem to be the tools to use. 
Reasonable implementations of the backward BTS reduction with IC3, of antichain, and of basic NFA should also be the baseline of comparisons.

\mona and \brics, based on DFA, as well as \jaltimpact focused on data words rather then on pure regular properties, do no reach the performance of the best tools. Also \bisim did not confirm the power of up-to algorithms. SMT-solvers, \zthree especially, are competitive, but cannot be considered the top of state of the art. 

Generally, the particular kind and source of benchmark is a decisive factor influencing the performance of tools, as especially visible on the parametric benchmark.

\paragraph{Threads to validity.}
Our results must be taken with a grain of salt as the experiment contains an inherent room for error. 
Although we tried to be as fair as possible, not knowing every tool intimately, the conversions between formats and kinds of automata, discussed at the start of \cref{sec:comparison}, might have introduced biases into the experiment.  
Tools are written in different languages and some have parameters which we might have used in sub-optimal way (we use the tools in their default settings), or, in the case of libraries, we could have used a sub-optimal combination of functions. 
We also did not measure memory peaks, which could be especially interesting e.g. in when the tools are deployed on a cloud.   
We are, however, confident that our main conclusions are well justified and the experiment gives a good overall picture. The entire experiment is available for anyone to challenge or improve upon \cite{experimentweb}.

\bibliography{literature}

\begin{thebibliography}{10}
\providecommand{\url}[1]{\texttt{#1}}
\providecommand{\urlprefix}{URL }
\providecommand{\doi}[1]{https://doi.org/#1}

\bibitem{experimentweb}
Experiment replication package and additional material,
  \url{https://www.fit.vutbr.cz/research/groups/verifit/tools/afa-comparison/}

\bibitem{jaltimpact}
Jaltimpact, \url{https://github.com/cathiec/JAltImpact}

\bibitem{Trau}
Abdulla, P.A., Atig, M.F., Chen, Y., Diep, B.P., Hol{\'{\i}}k, L., Rezine, A.,
  R{\"{u}}mmer, P.: Trau: {SMT} solver for string constraints. In: 2018 Formal
  Methods in Computer Aided Design, {FMCAD} 2018, Austin, TX, USA, October 30 -
  November 2, 2018. pp.~1--5. {IEEE} (2018). \doi{10.23919/FMCAD.2018.8602997},
  \url{https://doi.org/10.23919/FMCAD.2018.8602997}

\bibitem{ChainFree}
Abdulla, P.A., Atig, M.F., Diep, B.P., Hol{\'{\i}}k, L., Janku, P.: Chain-free
  string constraints. In: Automated Technology for Verification and Analysis -
  17th International Symposium, {ATVA} 2019, Taipei, Taiwan, October 28-31,
  2019, Proceedings. Lecture Notes in Computer Science, vol. 11781, pp.
  277--293. Springer (2019). \doi{10.1007/978-3-030-31784-3\_16},
  \url{https://doi.org/10.1007/978-3-030-31784-3\_16}

\bibitem{abdulla_when_2010}
Abdulla, P.A., Chen, Y.F., Hol{\'i}k, L., Mayr, R., Vojnar, T.: When simulation
  meets antichains. In: TACAS'10. LNCS, vol.~6015, pp. 158--174. Springer
  (2010)

\bibitem{brics}
et~al, A.M.: Brics automata library, \url{https://www.brics.dk/automaton/}

\bibitem{cvc5}
Barbosa, H., Barrett, C., Brain, M., Kremer, G., Lachnitt, H., Mann, M.,
  Mohamed, A., Mohamed, M., Niemetz, A., N{\"o}tzli, A., Ozdemir, A., Preiner,
  M., Reynolds, A., Sheng, Y., Tinelli, C., Zohar, Y.: cvc5: A versatile and
  industrial-strength smt solver. In: Tools and Algorithms for the Construction
  and Analysis of Systems. pp. 415--442. Springer International Publishing,
  Cham (2022)

\bibitem{BarFT-SMTLIB}
Barrett, C., Fontaine, P., Tinelli, C.: {The Satisfiability Modulo Theories
  Library ({SMT-LIB})}. {\tt www.SMT-LIB.org} (2016)

\bibitem{Z3str4}
{Berzish, Murphy}: Z3str4: A Solver for Theories over Strings. Ph.D. thesis
  (2021), \url{http://hdl.handle.net/10012/17102}

\bibitem{noodler23}
Blahoudek, F., Chen, Y.F., Chocholat{\'y}, D., Havlena, V., Hol{\'i}k, L.,
  Leng{\'a}l, O., S{\'i}{\v{c}}, J.: Word equations in synergy with regular
  constraints. In: Chechik, M., Katoen, J.P., Leucker, M. (eds.) Formal
  Methods. pp. 403--423. Springer International Publishing, Cham (2023)

\bibitem{bonchi_checking_2013}
Bonchi, F., Pous, D.: Checking {NFA} equivalence with bisimulations up to
  congruence. In: {POPL}'13. pp. 457--468. {{ACM} Trans. Comput. Log.} (2013)

\bibitem{bouajjani-antichain-08}
Bouajjani, A., Habermehl, P., Hol{\'i}k, L., Touili, T., Vojnar, T.:
  Antichain-based universality and inclusion testing over nondeterministic
  finite tree automata. In: Implementation and Applications of Automata. pp.
  57--67. Springer Berlin Heidelberg, Berlin, Heidelberg (2008)

\bibitem{BoudetC96}
Boudet, A., Comon, H.: {Diophantine} equations, {Presburger} arithmetic and
  finite automata. In: Kirchner, H. (ed.) Trees in Algebra and Programming -
  CAAP'96, 21st International Colloquium, Link{\"{o}}ping, Sweden, April,
  22-24, 1996, Proceedings. Lecture Notes in Computer Science, vol.~1059, pp.
  30--43. Springer (1996). \doi{10.1007/3-540-61064-2\_27},
  \url{https://doi.org/10.1007/3-540-61064-2\_27}

\bibitem{ltl:boeing}
Bozzano, M., Cimatti, A., Fernandes~Pires, A., Jones, D., Kimberly, G., Petri,
  T., Robinson, R., Tonetta, S.: Formal design and safety analysis of {AIR6110}
  wheel brake system. In: Kroening, D., P{\u{a}}s{\u{a}}reanu, C.S. (eds.)
  Computer Aided Verification. pp. 518--535. Springer International Publishing,
  Cham (2015)

\bibitem{bradley_safety_2007}
Bradley, A.R., Manna, Z.: Checking safety by inductive generalization of
  counterexamples to induction. In: {FMCAD}'07. pp. 173--180. IEEE Computer
  Society (2007)

\bibitem{ic3ref}
Bradley, A.: Ic3 reference implementation: a short, simple, fairly competitive
  implementation of ic3 (2015), \url{https://github.com/arbrad/IC3ref}

\bibitem{abc}
Brayton, R., Mishchenko, A.: Abc: An academic industrial-strength verification
  tool. In: CAV'10. pp. 24--40. Springer (2010)

\bibitem{Brzozowski1962CanonicalRE}
Brzozowski, J.A.: Canonical regular expressions and minimal state graphs for
  definite events. In: Proc. Symposium of Mathematical Theory of Automata. pp.
  529--561 (1962)

\bibitem{Buchi62}
B{\"u}chi, J.R.: On a decision method in restricted second order arithmetic.
  In: Proc. of International Congress on Logic, Method, and Philosophy of
  Science 1960. Stanford Univ. Press, Stanford (1962)

\bibitem{nuxmv}
Cavada, R., Cimatti, A., Dorigatti, M., Griggio, A., Mariotti, A., Micheli, A.,
  Mover, S., Roveri, M., Tonetta, S.: The nuxmv symbolic model checker. In:
  Computer Aided Verification. pp. 334--342. Springer International Publishing,
  Cham (2014)

\bibitem{Cece17}
C{\'{e}}c{\'{e}}, G.: Foundation for a series of efficient simulation
  algorithms. In: 32nd Annual {ACM/IEEE} Symposium on Logic in Computer
  Science, {LICS} 2017, Reykjavik, Iceland, June 20-23, 2017. pp. 1--12. {IEEE}
  Computer Society (2017). \doi{10.1109/LICS.2017.8005069},
  \url{https://doi.org/10.1109/LICS.2017.8005069}

\bibitem{AnthonyReplaceAll2018}
Chen, T., Chen, Y., Hague, M., Lin, A.W., Wu, Z.: What is decidable about
  string constraints with the replaceall function. Proc. {ACM} Program. Lang.
  \textbf{2}({POPL}),  3:1--3:29 (2018). \doi{10.1145/3158091},
  \url{https://doi.org/10.1145/3158091}

\bibitem{AnthonyComplex2019}
Chen, T., Hague, M., Lin, A.W., R{\"{u}}mmer, P., Wu, Z.: Decision procedures
  for path feasibility of string-manipulating programs with complex operations.
  Proc. {ACM} Program. Lang.  \textbf{3}({POPL}),  49:1--49:30 (2019).
  \doi{10.1145/3290362}, \url{https://doi.org/10.1145/3290362}

\bibitem{cox_mosca_19}
Cox, A.: {Model Checking Regular Expressions}.
  \textsc{url:}~\url{https://mosca19.github.io/slides/cox.pdf} (2019),
  presented at MOSCA'19

\bibitem{cox_paper_17}
Cox, A., Leasure, J.: Model checking regular language constraints. CoRR
  \textbf{abs/1708.09073} (2017)

\bibitem{lorisjava}
D'Anthoni, L.: symbolicautomata,
  \url{https://github.com/lorisdanto/symbolicautomata}

\bibitem{dantoni_afa_2016}
D'Antoni, L., Kincaid, Z., Wang, F.: A symbolic decision procedure for symbolic
  alternating finite automata. Electronic Notes in Theoretical Computer Science
   \textbf{336},  79--99 (2018).
  \doi{https://doi.org/10.1016/j.entcs.2018.03.017},
  \url{https://www.sciencedirect.com/science/article/pii/S1571066118300203},
  the Thirty-third Conference on the Mathematical Foundations of Programming
  Semantics (MFPS XXXIII)

\bibitem{dantoni_taminimization_2016}
D'Antoni, L., argus Veanes: Minimization of symbolic tree automata. In:
  {LICS}'16. pp. 873--882. {{ACM} Trans. Comput. Log.} (2016)

\bibitem{margus_minimization}
D'Antoni, L., Veanes, M.: Minimization of symbolic automata. In: Proceedings of
  the 41st ACM SIGPLAN-SIGACT Symposium on Principles of Programming Languages.
  p. 541–553. POPL '14, Association for Computing Machinery, New York, NY,
  USA (2014). \doi{10.1145/2535838.2535849},
  \url{https://doi.org/10.1145/2535838.2535849}

\bibitem{margus_the_power17}
D'Antoni, L., Veanes, M.: The power of symbolic automata and transducers. In:
  Computer Aided Verification. pp. 47--67. Springer International Publishing,
  Cham (2017)

\bibitem{ltlf}
De~Giacomo, G., Vardi, M.Y.: Linear temporal logic and linear dynamic logic on
  finite traces. In: IJCAI'13. pp. 854--860. ACM (2013)

\bibitem{alaska}
De~Wulf, M., Doyen, L., Maquet, N., Raskin, J.F.: Alaska. In: Cha, S.S., Choi,
  J.Y., Kim, M., Lee, I., Viswanathan, M. (eds.) Automated Technology for
  Verification and Analysis. pp. 240--245. Springer Berlin Heidelberg, Berlin,
  Heidelberg (2008)

\bibitem{doyen-antichain-10}
Doyen, L., Raskin, J.: Antichain algorithms for finite automata. In: Tools and
  Algorithms for the Construction and Analysis of Systems, 16th International
  Conference, {TACAS} 2010, Held as Part of the Joint European Conferences on
  Theory and Practice of Software, {ETAPS} 2010, Paphos, Cyprus, March 20-28,
  2010. Proceedings. Lecture Notes in Computer Science, vol.~6015, pp. 2--22.
  Springer (2010). \doi{10.1007/978-3-642-12002-2\_2},
  \url{https://doi.org/10.1007/978-3-642-12002-2\_2}

\bibitem{minisat03}
E{\'{e}}n, N., S{\"{o}}rensson, N.: An extensible sat-solver. In: Theory and
  Applications of Satisfiability Testing, 6th International Conference, {SAT}
  2003. Santa Margherita Ligure, Italy, May 5-8, 2003 Selected Revised Papers.
  Lecture Notes in Computer Science, vol.~2919, pp. 502--518. Springer (2003).
  \doi{10.1007/978-3-540-24605-3\_37},
  \url{https://doi.org/10.1007/978-3-540-24605-3\_37}

\bibitem{chenfu_eqchecking_17}
Fu, C., Deng, Y., Jansen, D.N., Zhang, L.: On equivalence checking of
  nondeterministic finite automata. In: Dependable Software Engineering.
  Theories, Tools, and Applications - Third International Symposium, {SETTA}
  2017, Changsha, China, October 23-25, 2017, Proceedings. Lecture Notes in
  Computer Science, vol. 10606, pp. 216--231. Springer (2017).
  \doi{10.1007/978-3-319-69483-2\_13},
  \url{https://doi.org/10.1007/978-3-319-69483-2\_13}

\bibitem{chenfu_equchecking_17}
Fu, C., Deng, Y., Jansen, D.N., Zhang, L.: On equivalence checking of
  nondeterministic finite automata. In: Dependable Software Engineering.
  Theories, Tools, and Applications. pp. 216--231. Springer International
  Publishing, Cham (2017)

\bibitem{gange_unbounded_13}
Gange, G., Navas, J.A., Stuckey, P.J., S{\o}ndergaard, H., Schachte, P.:
  Unbounded model-checking with interpolation for regular language constraints.
  In: {TACAS} 2013. Lecture Notes in Computer Science, vol.~7795, pp. 277--291.
  Springer (2013). \doi{10.1007/978-3-642-36742-7\_20},
  \url{https://doi.org/10.1007/978-3-642-36742-7\_20}

\bibitem{ganty_fixed_2010}
Ganty, P., Maquet, N., Raskin, J.: Fixed point guided abstraction refinement
  for alternating automata. Theor. Comput. Sci.  \textbf{411}(38-39),
  3444--3459 (2010)

\bibitem{ltl:nasa}
Gario, M., Cimatti, A., Mattarei, C., Tonetta, S., Rozier, K.Y.: Model checking
  at scale: Automated air traffic control design space exploration. In:
  Chaudhuri, S., Farzan, A. (eds.) Computer Aided Verification. pp. 3--22.
  Springer International Publishing, Cham (2016)

\bibitem{ltl:lift}
Harding, A.: Symbolic strategy synthesis for games with {LTL} winning
  conditions. Ph.D. thesis, University of Birmingham (2005)

\bibitem{mona}
Henriksen, J.G., Jensen, J.L., J{\o}rgensen, M.E., Klarlund, N., Paige, R.,
  Rauhe, T., Sandholm, A.: Mona: {M}onadic second-order logic in practice. In:
  {TACAS} '95. LNCS, vol.~1019, pp. 89--110. Springer (1995)

\bibitem{HHK95}
Henzinger, M.R., Henzinger, T.A., Kopke, P.W.: Computing simulations on finite
  and infinite graphs. In: 36th Annual Symposium on Foundations of Computer
  Science, Milwaukee, Wisconsin, USA, 23-25 October 1995. pp. 453--462. {IEEE}
  Computer Society (1995). \doi{10.1109/SFCS.1995.492576},
  \url{https://doi.org/10.1109/SFCS.1995.492576}

\bibitem{hoder_pdr_2012}
Hoder, K., Bj{\o}rner, N.: Generalized property directed reachability. In:
  {SAT}'12. LNCS, vol.~7317, pp. 157--171. Springer (2012)

\bibitem{janku_string_2018}
Hol{\'{\i}}k, L., Jank{\r u}, P., Lin, A.W., R{\"{u}}mmer, P., Vojnar, T.:
  String constraints with concatenation and transducers solved efficiently.
  Proc. {ACM} Program. Lang.  \textbf{2}({POPL}) (2018)

\bibitem{symbsim18}
Hol{\'i}k, L., Leng{\'a}l, O., S{\'i}{\v{c}}, J., Veanes, M., Vojnar, T.:
  Simulation algorithms for symbolic automata. In: Lahiri, S.K., Wang, C.
  (eds.) Automated Technology for Verification and Analysis. pp. 109--125.
  Springer International Publishing, Cham (2018)

\bibitem{tree_inclusion_11}
Hol{\'{\i}}k, L., Leng{\'{a}}l, O., Sim{\'{a}}cek, J., Vojnar, T.: Efficient
  inclusion checking on explicit and semi-symbolic tree automata. In: Bultan,
  T., Hsiung, P. (eds.) Automated Technology for Verification and Analysis, 9th
  International Symposium, {ATVA} 2011, Taipei, Taiwan, October 11-14, 2011.
  Proceedings. Lecture Notes in Computer Science, vol.~6996, pp. 243--258.
  Springer (2011). \doi{10.1007/978-3-642-24372-1\_18},
  \url{https://doi.org/10.1007/978-3-642-24372-1\_18}

\bibitem{lukasjirisimulation}
Hol\'{i}k, L., \v{S}im\'{a}\v{c}ek, J.: Optimizing an lts-simulation algorithm.
  Computing and Informatics  \textbf{2010}(7),  1337--1348 (2010),
  \url{https://www.fit.vut.cz/research/publication/9733}

\bibitem{dprle}
Hooimeijer, P., Weimer, W.: A decision procedure for subset constraints over
  regular languages. In: Proceedings of the 30th ACM SIGPLAN Conference on
  Programming Language Design and Implementation. p. 188–198. PLDI '09,
  Association for Computing Machinery, New York, NY, USA (2009).
  \doi{10.1145/1542476.1542498}, \url{https://doi.org/10.1145/1542476.1542498}

\bibitem{hopcroft_71}
Hopcroft, J.E.: An n log n algorithm for minimizing states in a finite
  automaton. Tech. rep., Stanford, CA, USA (1971)

\bibitem{HUFFMAN1954}
Huffman, D.: The synthesis of sequential switching circuits. Journal of the
  Franklin Institute  \textbf{257}(3),  161--190 (1954).
  \doi{https://doi.org/10.1016/0016-0032(54)90574-8},
  \url{https://www.sciencedirect.com/science/article/pii/0016003254905748}

\bibitem{Ilie2004}
Ilie, L., Navarro, G., Yu, S.: On NFA Reductions, pp. 112--124. Springer Berlin
  Heidelberg, Berlin, Heidelberg (2004). \doi{10.1007/978-3-540-27812-2_11},
  \url{https://doi.org/10.1007/978-3-540-27812-2_11}

\bibitem{radu18}
Iosif, R., Xu, X.: Abstraction refinement for emptiness checking of alternating
  data automata. In: TACAS'18. pp. 93--111. Springer (2018)

\bibitem{vata}
Leng{\'{a}}l, O., Sim{\'{a}}cek, J., Vojnar, T.: {VATA:} {A} library for
  efficient manipulation of non-deterministic tree automata. In: Tools and
  Algorithms for the Construction and Analysis of Systems - 18th International
  Conference, {TACAS} 2012, Held as Part of the European Joint Conferences on
  Theory and Practice of Software, {ETAPS} 2012, Tallinn, Estonia, March 24 -
  April 1, 2012. Proceedings. Lecture Notes in Computer Science, vol.~7214, pp.
  79--94. Springer (2012). \doi{10.1007/978-3-642-28756-5\_7},
  \url{https://doi.org/10.1007/978-3-642-28756-5\_7}

\bibitem{ltlf:multiplebenchmarks}
Li, J., Pu, G., Zhang, Y., Vardi, M.Y., Rozier, K.Y.: {SAT}-based explicit
  {LTLf} satisfiability checking. Artificial Intelligence  \textbf{289},
  103369 (2020). \doi{https://doi.org/10.1016/j.artint.2020.103369},
  \url{https://www.sciencedirect.com/science/article/pii/S0004370220301193}

\bibitem{libfa}
Lutterkort, D.: libfa, \url{https://augeas.net/libfa/}

\bibitem{macmillan-interpolation-03}
McMillan, K.L.: Interpolation and sat-based model checking. In: Computer Aided
  Verification. pp. 1--13. Springer Berlin Heidelberg, Berlin, Heidelberg
  (2003)

\bibitem{mcmillan_lazy_2006}
McMillan, K.L.: Lazy abstraction with interpolants. In: {CAV}'06. LNCS,
  vol.~4144, pp. 123--136. Springer (2006)

\bibitem{Moore1956}
Moore, E.F.: Gedanken-Experiments on Sequential Machines, pp. 129--154.
  Princeton University Press, Princeton (1956).
  \doi{doi:10.1515/9781400882618-006},
  \url{https://doi.org/10.1515/9781400882618-006}

\bibitem{z3}
de~Moura, L., Bj{\o}rner, N.: Z3: An efficient smt solver. In: Tools and
  Algorithms for the Construction and Analysis of Systems. pp. 337--340.
  Springer Berlin Heidelberg, Berlin, Heidelberg (2008)

\bibitem{cvc422}
N{\"o}tzli, A., Reynolds, A., Barbosa, H., Barrett, C., Tinelli, C.: Even
  faster conflicts and lazier reductions for string solvers. In: Computer
  Aided Verification. pp. 205--226. Springer International Publishing, Cham
  (2022)

\bibitem{piagetarjan_87}
Paige, R., Tarjan, R.E.: Three partition refinement algorithms. SIAM Journal on
  Computing  \textbf{16}(6),  973--989 (1987). \doi{10.1137/0216062},
  \url{https://doi.org/10.1137/0216062}

\bibitem{ranzato_efficient_2010}
Ranzato, F., Tapparo, F.: An efficient simulation algorithm based on abstract
  interpretation. Information and Computation  \textbf{208},  1--22 (2010)

\bibitem{regexlib}
RegExLib.com: {The Internet's first Regular Expression Library}.
  {\url{http://regexlib.com/}}

\bibitem{ltl:counter}
Rozier, K.Y., Vardi, M.Y.: {LTL} satisfiability checking. In: SPIN'07. pp.
  149--167. Springer (2007)

\bibitem{margus-derivatives-21}
Stanford, C., Veanes, M., Bj\o{}rner, N.: Symbolic boolean derivatives for
  efficiently solving extended regular expression constraints. In: Proceedings
  of the 42nd ACM SIGPLAN International Conference on Programming Language
  Design and Implementation. p. 620–635. PLDI 2021, Association for Computing
  Machinery, New York, NY, USA (2021). \doi{10.1145/3453483.3454066},
  \url{https://doi.org/10.1145/3453483.3454066}

\bibitem{margus_derivatives_21}
Stanford, C., Veanes, M., Bj{\o}rner, N.S.: Symbolic boolean derivatives for
  efficiently solving extended regular expression constraints. In: {PLDI} '21.
  pp. 620--635. {ACM} (2021). \doi{10.1145/3453483.3454066},
  \url{https://doi.org/10.1145/3453483.3454066}

\bibitem{tabakov-vardi_05}
Tabakov, D., Vardi, M.Y.: Experimental evaluation of classical automata
  constructions. In: Logic for Programming, Artificial Intelligence, and
  Reasoning. pp. 396--411. Springer Berlin Heidelberg, Berlin, Heidelberg
  (2005)

\bibitem{Valmari_10}
Valmari, A.: Simple bisimilarity minimization in o(m log n) time. Fundam.
  Informaticae  \textbf{105}(3),  319--339 (2010). \doi{10.3233/FI-2010-369},
  \url{https://doi.org/10.3233/FI-2010-369}

\bibitem{vargovcik_21}
Vargov{\v{c}}{\'i}k, P., Hol{\'i}k, L.: Simplifying alternating automata for
  emptiness testing. In: Programming Languages and Systems. pp. 243--264.
  Springer International Publishing, Cham (2021)

\bibitem{automatadotnet}
Veanes, M.: \url{https://github.com/AutomataDotNet}

\bibitem{margus_rex_10}
Veanes, M., de~Halleux, P., Tillmann, N.: Rex: Symbolic regular expression
  explorer. In: Third International Conference on Software Testing,
  Verification and Validation, {ICST} 2010, Paris, France, April 7-9, 2010. pp.
  498--507. {IEEE} Computer Society (2010). \doi{10.1109/ICST.2010.15},
  \url{https://doi.org/10.1109/ICST.2010.15}

\bibitem{fangyu_circuit_16}
Wang, H., Tsai, T., Lin, C., Yu, F., Jiang, J.R.: String analysis via automata
  manipulation with logic circuit representation. In: {CAV'16}. LNCS,
  vol.~9779, pp. 241--260. Springer (2016)

\bibitem{WolperB95}
Wolper, P., Boigelot, B.: An automata-theoretic approach to {Presburger}
  arithmetic constraints (extended abstract). In: Mycroft, A. (ed.) Static
  Analysis, Second International Symposium, SAS'95, Glasgow, UK, September
  25-27, 1995, Proceedings. Lecture Notes in Computer Science, vol.~983, pp.
  21--32. Springer (1995). \doi{10.1007/3-540-60360-3\_30},
  \url{https://doi.org/10.1007/3-540-60360-3\_30}

\bibitem{dewulf_antichains_2006}
Wulf, M.D., Doyen, L., Henzinger, T.A., Raskin, J.: Antichains: {A} new
  algorithm for checking universality of finite automata. In: {CAV}'06. LNCS,
  vol.~4144, pp. 17--30. Springer (2006)

\end{thebibliography}

\newpage
\eject
\appendix
\section{One to one comparison of the best tools}
\begin{figure}[h!]
  \centering
  \includegraphics[width=\textwidth]{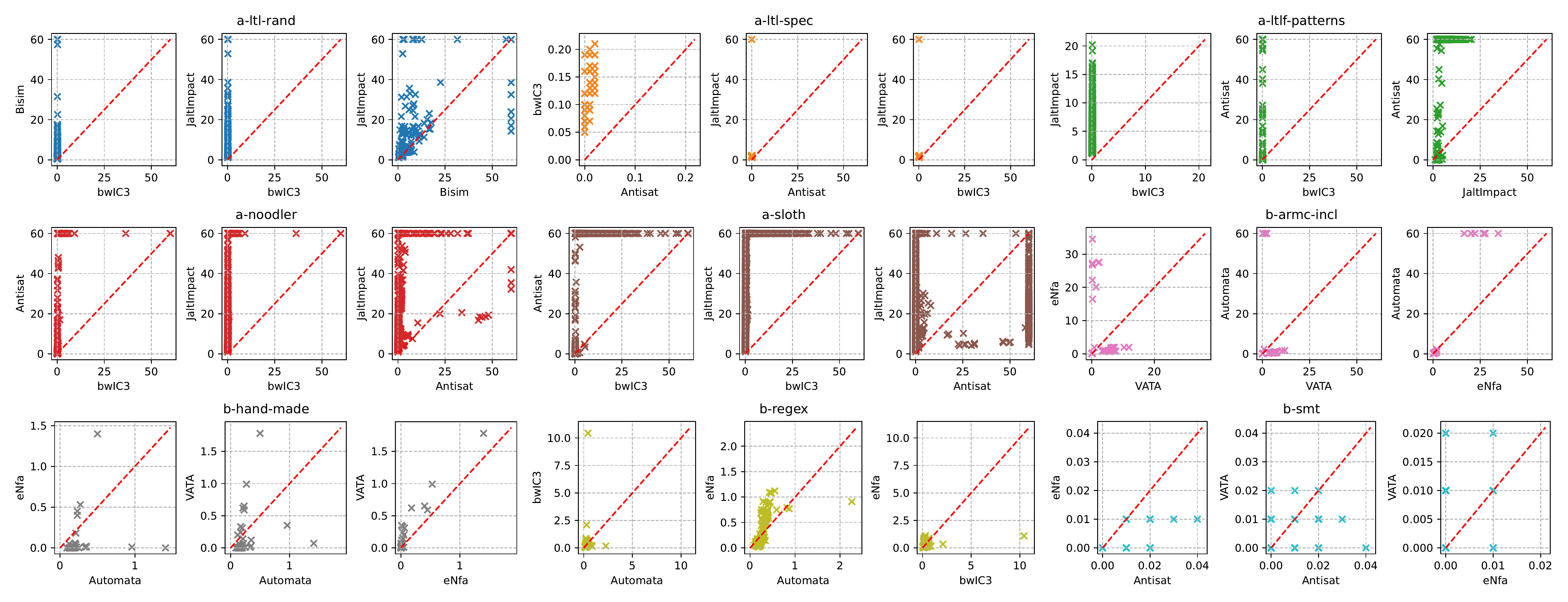}
        %\vspace{-10.0mm}
  \caption{Scatter-plots comparing the best tools one to one, by benchmark.}
  \label{fig:scatter}
\end{figure}

Further, for each benchmark shown on \cref{fig:cactus}, we have selected three best-performing tools
based on the rate of timeouts or their average times. We show their
comparison in Fig.~\ref{fig:scatter} in the form of scatter-plots. The triplets are distinguished by the color of the points and the position of each triplet corresponds to the position of the benchmark in \cref{fig:cactus}.

%\todo{research questions}

\end{document}